\documentclass[10pt,prd]{revtex4}
\usepackage{amssymb,epsfig}

\usepackage{amsmath}
\usepackage{graphicx,color}
\usepackage{amssymb,amsthm,subfigure}

\renewcommand{\thefootnote}{\fnsymbol{footnote}}
\def\half{\frac{1}{2}}\newcommand{\nc}{\newcommand}
\nc{\p}{\prime}
\nc{\figcaption}[1]{\def\@captype{figure}\caption{\scriptsize{#1}}}
\nc{\tblcaption}[1]{\def\@captype{table}\caption{\scriptsize{#1}}}

\nc{\be}[1]{\begin{equation} \mbox{$\label{#1}$}}
\nc{\ee}{\end{equation}}
\nc{\bea}[1]{\begin{eqnarray} \mbox{$\label{#1}$}}
\nc{\eea}{\end{eqnarray}}
\nc{\abs}[1]{\left|#1\right|}
\nc{\set}[1]{\left\{#1\right\}}
\nc{\bset}[1]{\left(#1\right)} \nc{\sbset}[1]{\left[#1\right]}
\nc{\eq}[1]{\mbox{Eq.~(\ref{#1})}}
\nc{\eqs}[2]{\mbox{Eqs.~(\ref{#1}, \ref{#2})}}
\nc{\fig}[1]{\mbox{Fig.~\ref{#1}}}
\nc{\figs}[2]{\mbox{Figs.~\ref{#1}, \ref{#2}}}
\nc{\tbl}[1]{\mbox{Table:~\ref{#1}}}
\nc{\Uo}{U_\omega} \nc{\So}{S_\omega}
\nc{\order}[1]{\mathcal{O}(#1)}
\nc{\eo}{\epsilon_\omega}
\nc{\mo}{m_\omega}
\nc{\mS}{\mathcal{S}}
\nc{\mU}{\mathcal{U}}
\nc{\tim}{\tilde{m}}
\nc{\tiom}{\tilde{\omega}}
\nc{\tisig}{\tilde{\sigma}}
\nc{\tir}{\tilde{r}}
\nc{\alom}{\kappa_\omega}
\nc{\lamom}{\lambda_\omega}

\begin{document}

\title{$Q$-balls in flat potentials}

\author{Edmund J. Copeland\footnote{ed.copeland@nottingham.ac.uk}  and
Mitsuo I. Tsumagari\footnote{ppxmt@nottingham.ac.uk} } 
\affiliation{\vspace{.5cm}\\  School of Physics and Astronomy, University of Nottingham, University Park, Nottingham NG7 2RD, UK}

\begin{abstract}
We study the classical and absolute stability of $Q$-balls in scalar field theories with flat potentials arising in both gravity-mediated  and gauge-mediated models. We show that the associated $Q$-matter formed in gravity-mediated potentials can be stable against decay into their own free particles as long as the coupling constant of the nonrenormalisable term is small, and that all of the possible three-dimensional $Q$-ball configurations are classically stable against linear fluctuations. Three-dimensional gauge-mediated $Q$-balls can be absolutely stable in the ``thin-wall-like'' limit, but are completely unstable in the ``thick-wall'' limit.
\end{abstract}

\vskip 1pc 

\maketitle


\renewcommand{\thefootnote}{\arabic{footnote}}
\setcounter{footnote}{0}

\section{Introduction}
$Q$-balls have recently attracted much attentions in cosmology \cite{Dvali:1997qv} and astrophysics \cite{Kusenko:1997vp, Cecchini:2008su, Takenaga:2006nr}. A $Q$-ball \cite{Coleman:1985ki} is a nontopological soliton \cite{Friedberg:1976me} whose stability is ensured by the existence of a continuous global charge $Q$ (for a review see \cite{Lee:1991ax, raj, Dine:2003ax, Enqvist:2003gh, Radu:2008pp} and references therein), and a number of scalar field theory models have been proposed to support the existence of nontopological solitons. They include polynomial models \cite{Coleman:1985ki}, Sine-Gordon models \cite{Arodz:2008jk}, parabolic-type models \cite{Theodorakis:2000bz}, confinement models \cite{Morris:1978ns, Simonov:1979rd, Mathieu:1987mr, Werle:1977hs}, two-field models \cite{Friedberg:1976me, McDonald:2001iv}, and flat models \cite{Dvali:1997qv}.

From a phenomenological point of view, the most interesting examples are probably the supersymmetric $Q$-balls arising within the framework of the Minimal Supersymmetric Standard Model (MSSM), which naturally contains a number of gauge invariant flat directions. Many of the flat directions can carry baryon (B) or/and lepton (L) number which is/are essential for Affleck-Dine (AD) baryogenesis \cite{Affleck:1984fy}. Following the AD mechanism, a complex scalar (AD) field acquires a large field value during a period of cosmic inflation and tends to form a homogeneous condensate, the AD condensate. In the presence of a negative pressure \cite{Lee:1994qb, Enqvist:1997si}, the condensate is unstable against spatial fluctuations so that it develops into nonlinear inhomogeneous lumps, namely $Q$-balls. The stationary properties and cosmological consequences of the $Q$-balls depend on how the Supersymmetry (SUSY) is broken in the hidden sector, transmitting to the observable sector through so-called messengers. In the gravity-mediated \cite{de Gouvea:1997tn} or gauge-mediated scenarios \cite{Dvali:1997qv}, the messengers correspond respectively either to supergravity fields or to some heavy particles charged under the gauge group of the standard model.

$Q$-balls can exist in scalar field potentials where SUSY is broken through effects in the supergravity hidden sector \cite{Nilles:1983ge}. These type of $Q$-balls can be unstable to decay into baryons and the lightest supersymmetric particle dark matter, such as neutralinos \cite{Fujii:2001xp}, gravitinos \cite{Ellis:1984eq, Seto:2005pj} and axinos \cite{Seto:2007ym}.  Recently, McDonald has argued that enhanced $Q$-ball decay in AD baryogenesis models can explain the 
observed positron and electron excesses detected by PAMELA, ATIC and PPB-BETS \cite{McDonald:2009cc}. By imposing an upper bound on the reheating temperature of the Universe after inflation, this mode of decay through $Q$-balls has been used to explain why the observed baryonic ($\Omega_{b}$) and dark matter ($\Omega_{DM}$) energy densities are so  similar \cite{Kusenko:1997si, Enqvist:1998en}, i.e. $\Omega_{DM}/\Omega_{b}=5.65\pm0.58$ \cite{Spergel:2006hy}.

Scalar field potentials arising through gauge-mediated SUSY breaking \cite{de Gouvea:1997tn} tend to be extremely flat. Using one of the MSSM flat directions, namely the $QdL$  direction (where $Q$ and $d$ correspond to squark fields and $L$ to a slepton field), which has a nonzero value of $B-L$ and therefore does not spoil AD baryogenesis via the sphaleron processes that violate $B+L$ \cite{Enqvist:1998en}, Shoemaker and Kusenko recently explored the minimum energy configuration for baryo-leptonic $Q$-balls, whose scalar field consists of both squarks and sleptons \cite{Shoemaker:2008gs}. It had been assumed to that point that the lowest energy state of the scalar field corresponds to being exactly the flat direction; however in \cite{Shoemaker:2008gs}, the authors showed that the lowest energy state lies slightly away from the flat directions, and that the relic $Q$-balls, which are stable against decay into both protons/neutrons (baryons) and neutrinos/electrons (leptons) \cite{Cohen:1986ct}, may end up contributing to the energy density of dark matter \cite{Kusenko:1997si, Laine:1998rg}; thus, the $Q$-balls can provide the baryon-to-photon ratio \cite{Laine:1998rg}, i.e. $n_b/n_\gamma \simeq (4.7-6.5) \times 10^{-10}$ \cite{Fields:2008zz}  where $n_b$ and $n_\gamma$ are respectively the baryon and photon number densities in the Universe.

In this paper we examine analytically and numerically the classical and absolute stability of $Q$-balls using flat potentials in the two specific models mentioned above. In order to study the possible existence of lower-dimensional $Q$-balls embedded in 3+1 dimensions, we will work in arbitrary spatial dimensions $D$; although of course the $D=3$ case is of more phenomenological interest. Previous work \cite{Enqvist:1997si, Enqvist:1998en, Multamaki:1999an} on the gravity-mediated potential has used either a steplike or Gaussian ansatz to study the analytical properties of the thin and thick-wall $Q$-balls. Introducing more physically motivated ans\"{a}tze, we will show that the thin-wall $Q$-balls can be quantum mechanically stable against decay into their own free particle quanta, that both thin and thick-wall $Q$-ball solutions obtained are classically stable against linear fluctuations, and confirm that a Gaussian ansatz is a physically reasonable one for the thick-wall $Q$-ball. The one-dimensional $Q$-balls in the thin-wall limit are excluded from our analytical framework. The literature on $Q$-balls with gauge-mediated potentials has tended to use a test profile in approximately flat potentials. We will present an exact profile for a generalised gauge-mediated flat potential, and show that we naturally recover results previously published in \cite{de Gouvea:1997tn, Enqvist:1998en, Laine:1998rg}.

The rest of this paper is organised as follows. In Sec. \ref{basics} we briefly review the important $Q$-ball properties that were established in \cite{Tsumagari:2008bv}. Section \ref{gravity-mediated} provides a detailed analyses for gravity-mediated potentials, and in Sec. \ref{gauge-mediated} we investigate the case of a generalised gauge-mediated potential. We confirm the validity of our analytical approximations with complete numerical $Q$-ball solutions in Sec. \ref{numerics} before summarising in Sec. \ref{conc}. Two appendices are included. In Appendix \ref{exactsol}, we obtain an exact solution for the case of a logarithmic potential, and in Appendix \ref{appxthick}, we confirm that the adoption of a Gaussian ansatz is appropriate for the thick-wall $Q$-ball found in the gravity-mediated potentials.

\section{The Basics}
\label{basics}

Here, we review the basic properties of $Q$-balls as described in \cite{Tsumagari:2008bv} and introduce a powerful technique that enables us to find the charge $Q$ and energy $E_Q$ of the $Q$-ball as well as the condition for its stability, and characteristic slope $\gamma(\omega) \equiv E_Q/\omega Q$ where $\omega$ is defined through the $Q$-ball ansatz, which is given by decomposing a complex scalar field $\phi$ into $\phi=\sigma(r)e^{i\omega t}$.  $\sigma$ is a real scalar field, $r$ is a radial coordinate, and therefore $\omega$ is a rotational frequency in the U(1) internal space. By scaling the radius $r$ of the $Q$-ball ansatz, which minimises $E_Q$,  we can find the characteristic slopes in terms of the ratio between the surface energy $\mS$ and the potential energy $\mU$ of the $Q$-ball. When the characteristic slope, $\gamma$, is independent of $\omega$, we obtain the relation: $E_Q \propto Q^{1/\gamma}$. In general the charge, energy and Euclidean action $\So$ are given by 
\be{euc}
Q=\omega \int_{V_D}\sigma^2; \hspace*{10pt}  \So=\int_{V_D} \left(\half \sigma^{\p 2} + \Uo\right); \hspace*{10pt} E_Q=\omega Q + \So,
\ee
where our metric is $ds^2=-dt^2 + h_{ij}dx^i dx^j$, the determinant of the spherically symmetric spatial metric $h_{ij}$ is defined by $h\equiv det(h_{ij})$, and we have used the following notation:
$\int_{V_D}\equiv \int d^D x \sqrt{h} =\Omega_{D-1}\int^\infty_0 dr\ r^{D-1}$, $\Omega_{D-1}\equiv \frac{2\pi^{D/2}}{\Gamma(D/2)}$, $\sigma^\p\equiv \frac{d\sigma}{dr}$, and $D$ is the number of spatial dimensions. Without loss of generality, we can take positive values of $\omega$ and $Q$.  By defining the effective potential $\Uo$ of a potential $U(\sigma)$
\be{Uomega}
\Uo\equiv U - \half \omega^2 \sigma^2,
\ee
the $Q$-ball equation is
\be{QBeq}
\sigma^{\p\p}+\frac{D-1}{r}\sigma^{\p}=\frac{d\Uo}{d\sigma},
\ee
where $\sigma(r)$ is a monotonically decreasing function in terms of $r$.
Given a potential $U(\sigma)$, which has a global minimum at $\sigma=0$, it is possible to show that $Q$-balls exist within the restricted range of $\omega$ \cite{Coleman:1985ki}:
\be{QBexist}
\omega_- \le \omega < \omega_+,
\ee
where we have defined the lower limit $\omega^2_-\equiv \left.\frac{2U}{\sigma^2}\right|_{\sigma_+(\omega_-)} \ge 0$, $\sigma_+(\omega)$ is the nonzero value of $\sigma$ where $U_{\omega}(\sigma_+(\omega))$ is minimised (see \fig{fig:gravpot}), and the upper limit $\omega^2_+ \equiv \left.\frac{d^2 U}{d\sigma^2}\right|_{\sigma=0}$. The existence condition \eq{QBexist} restricts the allowed form of the potential $U$, which implies that the potential should grow less quickly than the quadratic term (i.e. mass term) for small values of $\sigma$. The case $\omega_-=0$ corresponds to degenerate vacua potentials (DVPs), whilst $\omega_-\neq 0$ has nondegenerate vacua (NDVPs). In \cite{Tsumagari:2008bv} we examined the case of polynomial potentials and restricted ourselves to the case of $\omega^2_+=m^2$ where $m$ is a bare mass in the potentials. In this paper we extend our analysis allowing us to investigate the case  $\omega^2_+ \gg m^2$, needed since the potentials include one-loop corrections to the bare mass $m$. Here, the potential which we will consider in the gravity-mediated models, is $U=U_{grav}+U_{NR}$ where $U_{NR}$ is a nonrenormalisable term (to be discussed below), and 
\be{Ugrav}
U_{grav}\equiv \half m^2 \sigma^2\bset{1+K\ln\bset{\frac{\sigma^2}{M^2}}}.
\ee
Here, $K$ is a constant factor arising from the one-loop correction and  $M$ is the renormalisation scale.
To proceed with analytical arguments, we consider the two limiting values of $\omega$ or $\sigma_0\equiv \sigma(r=0)$ which describe

\be{QBdef}
\begin{cases}
  \bullet \hspace*{5pt} \textrm{thin-wall $Q$-balls when}\; \omega\simeq \omega_-\; \textrm{or equivalently}\; \sigma_0 \sim \sigma_+(\omega),\\
  \bullet \hspace*{5pt} \textrm{thick-wall $Q$-balls when}\; \omega\simeq \omega_+\; \textrm{or equivalently}\; \sigma_0 \simeq \sigma_-(\omega).
\end{cases}
\ee
Note, this limit doe not imply that a thick-wall $Q$-ball has to have a large thickness that is comparable to the size of the core size. For the extreme thin-wall limit, $\omega = \omega_-$, thin-wall $Q$-balls satisfy $\frac{E_Q}{Q}=\gamma(\omega_-) \omega_-$. In particular, Coleman demonstrated that a steplike profile for $Q$-balls, which generally exist for $\omega_-\neq 0$, satisfies $\gamma=1$, which implies that the charge $Q$ and energy $E_Q$ are proportional to the volume, and he called this $Q$-matter \cite{Coleman:1985ki}. For absolutely stable $Q$-balls, the energy per unit charge is smaller than the rest mass $m$ for the field $\phi$,
\be{QBstb}
\frac{E_Q}{Q} < m.
\ee
Thus the $Q$-ball satisfying \eq{QBstb} is stable against free-particle decays because the $Q$-ball energy $E_Q$ is less than a collection of $Q$ free-particles of total energy $E_{free} = mQ$. If the $Q$-ball has decay channels into other fundamental scalar particles that have the lowest mass $m_{min}$, we need to replace $m$ by $m_{min}$ in the absolute stability condition \eq{QBstb}. In the opposite limit $\omega\simeq \omega_+$, the $Q$-ball energy approaches the free particle energy, $E_Q \to mQ$. For later convenience, we define two positive definite quantities, $\eo$ and $\mo$ by
\begin{eqnarray}
\nonumber \eo&\equiv& -U_\omega(\sigma_+(\omega))=\half \omega^2 \sigma^2_+(\omega)- U(\sigma_+(\omega)),\\
\label{infqt} &\simeq& \half \bset{\omega^2-\omega^2_-}\sigma^2_+,\\ 
\label{mo} m^2_\omega&\equiv& m^2-\omega^2
\end{eqnarray}
which can be infinitesimally small for either thin- or thick-wall limits. By assuming $\sigma_+(\omega)\simeq \sigma_+(\omega_-) \equiv \sigma_+$ in the thin-wall limit, we immediately obtain the second line in \eq{infqt}. Notice that this assumption was implicitly imposed in our previous thin-wall analysis \cite{Tsumagari:2008bv}. While this is fine for the gravity-mediated case, with Gauge mediated potentials which are extremely flat, this implicit assumption cannot hold because $\sigma_+(\omega)$ does not exist. Therefore we will not use the variable $\epsilon_\omega$ for the case of the Gauge mediated potentials. 
Notice that the variable $m^2_\omega$ cannot be infinitesimally small when we consider the gravity-mediated case: $\omega^2_+ \not\sim m^2$.
A powerful tool we can make use of when calculating some of the physical $Q$-ball parameters is the Legendre relation \cite{Tsumagari:2008bv, Paccetti:2001uh}. For example the energy follows from 
\be{leg}
\So\to Q=-\left.\frac{d\So}{d\omega}\right|_{E_Q}\to E_Q=\omega Q + \So.
\ee
Assuming that $\gamma$ is not a function of $\omega$, we can compute the advertised characteristic slope,
\be{leg2}
\frac{E_Q}{\omega Q}=\gamma \to E_Q \propto Q^{1/\gamma}
\ee
where we have used another Legendre relation $\omega = \left.\frac{dE_Q}{dQ}\right|_{\So}$ in which we have fixed $\So$. If a $Q$-ball is classically stable, it satisfies 
\be{QBcls}
\frac{\omega}{Q}\frac{dQ}{d\omega}\le 0 \Leftrightarrow \frac{d}{d\omega}\bset{\frac{E_Q}{Q}}=-\frac{\So}{Q^2}\frac{dQ}{d\omega}\ge 0.
\ee
These classical stability conditions are equivalent to the fission condition, i.e. $\frac{d\omega}{dQ}\le 0$ in \cite{Tsumagari:2008bv} so that the charge $Q$ for classically stable $Q$-balls is a decreasing function in terms of $\omega$. By scaling a $Q$-ball solution with respect to the radius $r$, we also obtain the virial relation $D\mU=-(D-2)\mS+D\omega Q/2$ and the characteristic slope $\gamma(\omega)$,
\be{viri}
\gamma(\omega)=1+\bset{D-2+D\frac{\mU}{\mS}}^{-1}
\ee
once the ratio $\mS/\mU$ is given where $\mS\equiv \int_{V_D} \half \sigma^{\p 2}$ and $\mU\equiv \int_{V_D} U$ are the surface and potential energies, respectively. For $D\ge2$, we can see $\gamma(\omega)\ge 1$ because $\mS,\; \mU \ge 0$, which implies that $\So$ is positive definite for $D\ge2$, see \eq{euc}, whilst $\So$ is positive for $D=1$ only when $\mU \ge \mS$. It implies that we have to be careful to use the second relation of \eq{QBcls} for $D=1$ to evaluate the classical stability condition as we saw in the case of using the Gaussian ansatz, which is valid for $D=1$ for polynomial potentials \cite{Tsumagari:2008bv}. Our key results for $D\ge 2$ are
\be{virich}
    \gamma \simeq 
  \left\{
    \begin{array}{ll}
	1 \hspace*{10pt}  \textrm{for}\;  \mS \ll \mU,\\
	(2D-1)/2(D-1) \hspace*{10pt} \textrm{for}\; \mS \sim \mU,\\
	(D-1)/(D-2) \hspace*{10pt} \textrm{for}\; \mS \gg \mU.
    \end{array}
    \right.
\ee
The first case in \eq{virich} corresponds to the extreme case of thin and thick-wall $Q$-balls. Furthermore, in \cite{Tsumagari:2008bv}, we saw that for the extreme thin-wall $Q$-balls in DVPs, then there was a virialisation between $\mS$ and $\mU$, which corresponds to the second case in \eq{virich}. At present it is not known what kind of $Q$-ball potentials correspond to the third case; therefore, we will not be considering that case in the rest of our paper. Notice that in the case $\mS \gg \mU$ for $D=2$, we obtain the characteristic slope $\gamma\gg1$ from \eq{viri}. Similarly, for $D=1$, the characteristic slopes are obtained, i.e. $\gamma\simeq 1,\; \gg 1,\; \simeq 0$, respectively for $\mS\ll\mU,\; \mS \sim \mU,\; \mS \gg \mU$. We will use these $1D$ analytic results to interpret numerical results of one-dimensional $Q$-balls in the thin-wall limit. 

To end this section we note a nice duality that appears in \eqs{viri}{virich} between the two cases $ \mS \sim \mU$ and $ \mS \gg \mU$. In particular, for $ \mS \sim \mU$ in $D$ dimensions, the same result for $\gamma$ is obtained (to leading order) in $2\times D$ dimensions when  $ \mS \gg \mU$.

\section{Gravity-mediated potentials}
\label{gravity-mediated}

The MSSM consists of a number of flat directions where SUSY is not broken. Those flat directions are, however, lifted by gauge, gravity, and/or nonrenormalisable interactions. In what follows the gravity interaction is included perturbatively via the one-loop corrections for the bare mass $m$ in \eq{Ugrav} and the nonrenormalisable interactions ($U_{NR}$), which are suppressed by high energy scales such as the grand unified theory scale $M_U\sim 10^{16}$ GeV or Planck scale $m_{pl} \sim 10^{18}$ GeV.  Here, $m$ is of order of SUSY breaking scale which could be the gravitino mass $\sim m_{3/2}$, evaluated at the renormalisation scale $M$ \cite{Nilles:1983ge}. We note that, following the majority of work in this field, we will ignore A-term contributions ( U(1) violation terms), thermal effects \cite{Allahverdi:2000zd, Anisimov:2000wx} which come from the interactions between the AD field and the decay products of the inflaton, and the Hubble-induced terms which gives a negative mass-squared contribution during inflation. It is possible that their inclusion could well change the results of the following analysis.

The scalar potential we are considering at present is \cite{Enqvist:1997si, Nilles:1983ge}
\be{sugra}
U=U_{grav}+U_{NR}=\half m^2 \sigma^2\bset{1+K\ln \bset{\frac{\sigma^2}{M^2}}} + \frac{|\lambda|^2}{m^{n-4}_{pl}} \sigma^{n}
\ee
where we used \eq{Ugrav}, $K$ is a factor for the gaugino correction, which depends on the flat directions, and $M$ is the renormalisation scale. Also $\lambda$ is a dimensionless coupling constant, and $U_{NR}\equiv \frac{|\lambda|^2}{m^{n-4}_{pl}} \sigma^{n}$, where $n>2$. If the MSSM flat directions include a large top quark, $K$ can be positive and then $Q$-balls do not exist. For flat directions that do not have a large top quark component, we typical find $K \simeq -[0.01-0.1]$ \cite{Enqvist:1997si, Enqvist:2000gq}. The power $n$ of the nonrenormalisable term depends on the flat directions we are choosing along which we maintain R parity. As examples of the directions involving squarks, the $u^c d^c d^c$ direction has $n=6$, whilst the $u^c u^c d^c e^c$ direction requires $n=10$. A complete list of the MSSM flat directions can be found in Table 1 of \cite{Dine:1995kz}. Since the potential in \eq{sugra} for $K<0$ could satisfy the $Q$-ball existence condition in \eq{QBexist}, where $\omega_+ \gg m$, $Q$-balls naturally exist.

In the rest of this paper, we will focus on potentials of the form of \eq{sugra} for general $D(\ge 1)$ and $\omega$ and $n (>2)$ so that $M$ and $m_{pl}$ have the same mass dimension, $(D-1)/2$, as $\sigma$. It means that the parameters $M$ and $m_{pl}$ are only physical for $D=3$. For several cases of $n$ and $D$, the term $U_{NR}$ can be renormalisable, but we will generally call it the nonrenormalisable term for the future convenience. The readers should note that the potential \eq{sugra} has been derived only with $\mathcal{N}=1$ supergravity in $D=3$; therefore, the potential form could well be changed in other dimensions. Furthermore, the logarithmic correction breaks down for small $\sigma$ and the curvature of \eq{sugra} at $\sigma=0$ is finite due to the gaugino mass, which affects our thick-wall analysis and their dynamics. However, we concentrate our analysis on this potential form for arbitrary $D,\; n$ and any values of $\sigma$ for two main reasons. The first is that it contains a number of general semiclassical features expected of all the potentials, and the second is that it offers the opportunity to consider the lower-dimensional $Q$-balls embedded in $D=3$.

In Appendix \ref{exactsol}, we obtain the exact solution of \eq{QBeq} with the potential $U=U_{grav}$; however, exact solutions of the general potential $U$ in \eq{sugra} are fully nonlinear and can be obtained only numerically. Therefore, we will analytically examine the approximate solutions in both the thin and thick-wall limits.  Before doing so, we shall begin by imposing a restriction on $\lambda$ in \eq{sugra} in order to obtain stable $Q$-matter in NDVPs. With the further restrictions on $\lambda$ and $|K|$, we can proceed with our analytical arguments, and we will finally obtain the asymptotic $Q$-ball profile for large $r$ which will be used in the numerical section, Sec. \ref{numerics}.
 
\subsection{The existence of absolutely stable $Q$-matter}\label{exstQmat}

As we have seen, the first restriction on the gravity-mediated potential \eq{sugra} which will  allow for the existence of a $Q$-ball solution \eq{QBexist} is $K<0$. However, given values for $m,\; m_{pl},\; M$, $n$, and $K$ in \eq{sugra}, we need to restrict the allowed values of the parameter, $\lambda$ in the potential in order to ensure we obtain  absolutely stable $Q$-matter. Notice that $Q$-matter exists in NDVPs, whilst the extreme thin-wall $Q$-balls in DVPs, which will not be $Q$-matter as it will turn out, may exist with the lowest possible limit of $\lambda$.

By using the definitions of $\omega_-$ and $\sigma_+$, namely,
$\omega^2_-\equiv \left.\frac{2U}{\sigma^2}\right|_{\sigma_+}$ and $\left.\frac{d U_{\omega_-}}{d\sigma}\right|_{\sigma_+}=0$, we shall find the range of values of $\lambda$ for which absolutely stable $Q$-matter solutions exist. Moreover, we will obtain the curvature $\mu$, which is proportional to $\abs{K}$, of the effective potential $\Uo$ at $\sigma_+$.

\begin{figure}[!ht]
  \begin{center}
	\includegraphics[angle=-90, scale=0.31]{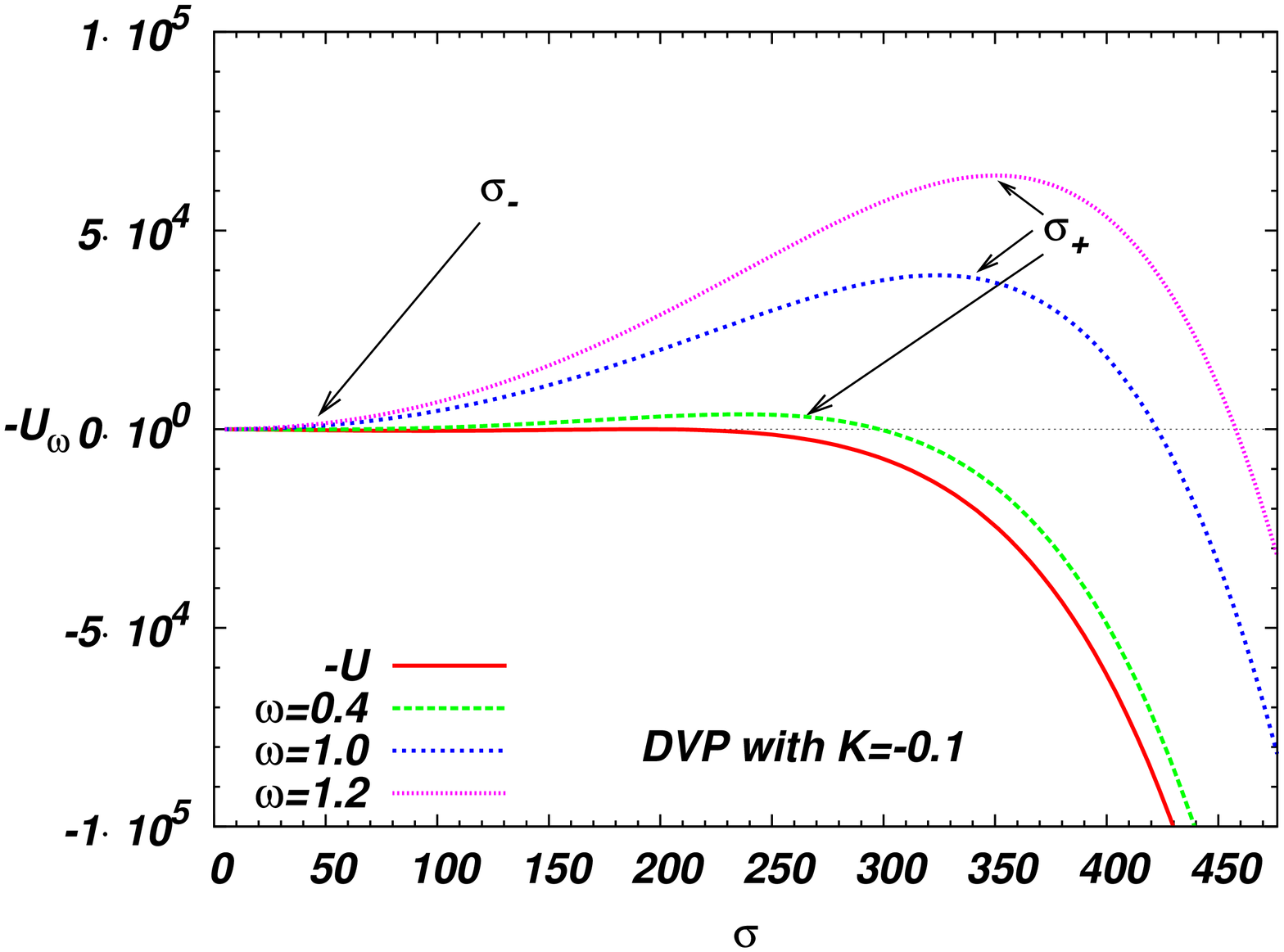}
	\includegraphics[angle=-90, scale=0.31]{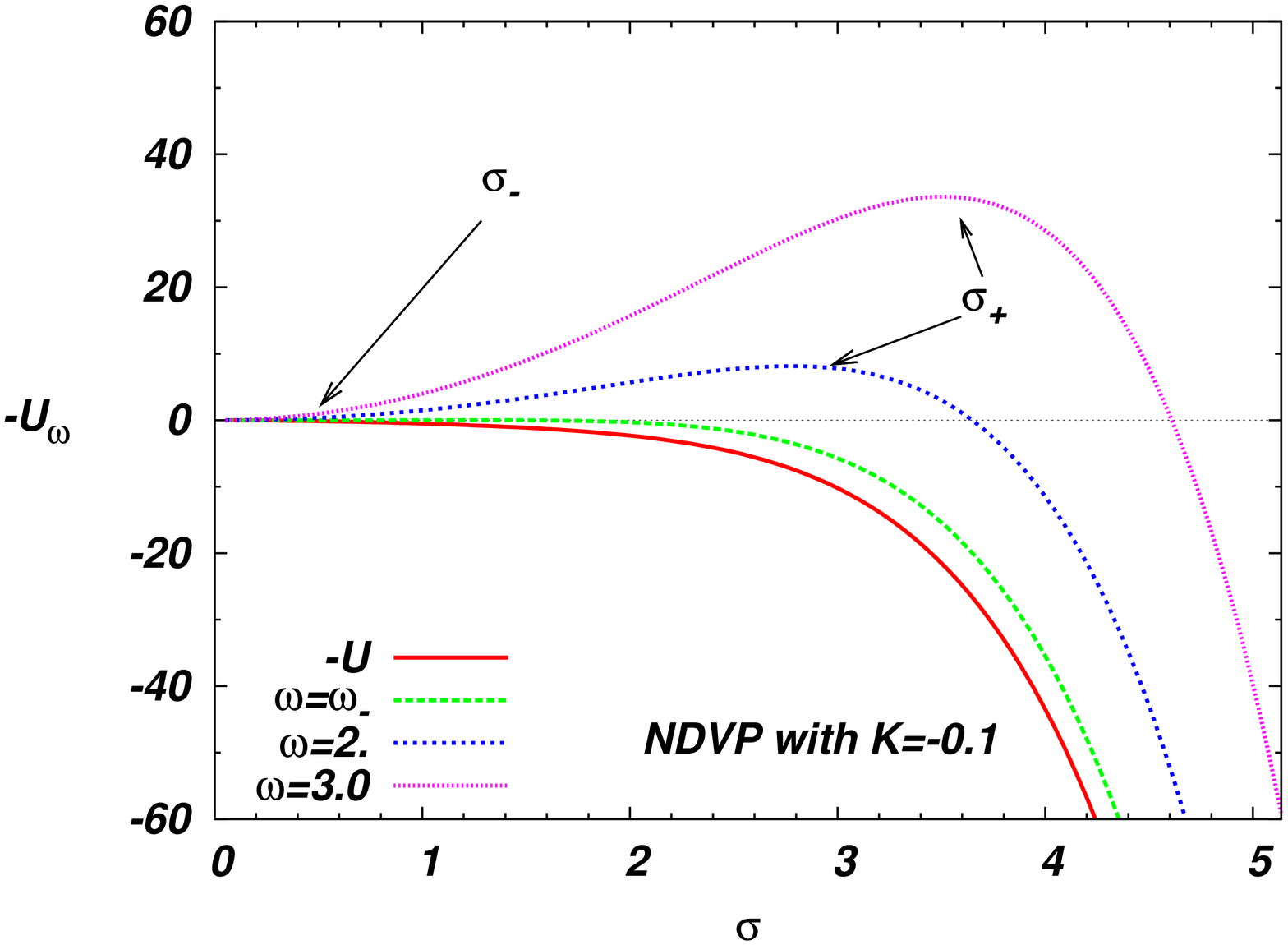}
  \end{center}
  \caption{Parameters $\sigma_{\pm}(\omega)$ for a potential of the form $U(\sigma)=\half \sigma^2\bset{1-|K|\ln\sigma^2} + \beta^2 \sigma^6$ (effective potential $\Uo=U-\half \omega^2\sigma^2$) with $|K|=0.1$. The left hand figure corresponds to the case of a DVP with $\beta^2=\frac{|K|e^{-1}}{4}\exp\bset{-\frac{2}{|K|}} \sim 1.90 \times 10^{-11}$, whilst the right hand side is the NDVP with $\beta^2=\frac{|K|e^{-1}}{4} \sim 9.20 \times 10^{-3}$, see \eq{rbeta}. The coloured lines in each plot correspond to different values of $\omega$. The variable
$\sigma_+(\omega)$ is defined as the maximum of the inverse effective potential $-\Uo$ whereas $\sigma_-(\omega)$ corresponds to $-\Uo(\sigma_-(\omega))=0$ for $\sigma_-(\omega)\neq 0$. Recalling $\omega_-=0$ in DVP, the DVP has degenerate vacua at $\sigma_+(0)=e^{1/4}\exp\bset{\frac{1}{2|K|}}\sim 1.91\times 10^2$ (red-solid line), whilst the NDVP does not. The inverse effective potential $-\Uo$ with $\omega_-=1$ in NDVP (green-dashed line), however, has degenerate vacua at $\sigma_+(\omega_-)=e^{1/4}\sim 1.28$, see the first relation in \eq{tisigom}. For the lower limit $\omega\simeq \omega_-$ (green-dashed lines), we could see $\sigma_+=e^{1/4}$, whilst the purple dotted-dashed lines show $\sigma_-(\omega) \to 0$ near the thick-wall like limit $\omega =3.0\sim \omega_+$ where $\omega_+ \gg 1$.}
  \label{fig:gravpot}
\end{figure}

The effective potential for \eq{sugra} can be rewritten in terms of new dimensionless variables $\tisig=\sigma/M,\; \tiom=\omega/m$, and \be{beta2}
\beta^2=\frac{|\lambda|^2 M^{n-2}}{m^{n-4}_{pl} m^2}>0,
\ee
as
\be{repsugra}
 U_{\tilde{\omega}}=\half M^2 m^2 \tisig^2\bset{1-\tiom^2 - 2|K|\ln\tisig} + M^2 m^2 \beta^2 \tisig^n.
\ee
After some simple algebra and introducing $\tiom^2_-\equiv \frac{2U}{\tisig^2}|_{\tisig_+}$ and $\left.\frac{d U_{\tiom_-}}{d\tisig}\right|_{\tisig_+}=0$, we obtain 
\be{tisigom}
\tisig_+=\bset{\frac{|K|}{(n-2)\beta^2}}^{\frac{1}{n-2}},\hspace*{10pt} \tiom^2_-=\frac{1}{n-2}\sbset{n-2+2|K|-2|K|\ln\bset{\frac{|K|}{(n-2)\beta^2}}}.
\ee
Notice that $\tiom^2_-=0$ corresponds to DVPs where $Q$-matter solutions do not exist \cite{Tsumagari:2008bv}, whilst the extreme thin-wall $Q$-balls do exist and are absolutely stable as we will see. In NDVPs, $Q$-matter solutions exist and are absolutely stable when $0<\tiom^2_-< 1$, see \eq{QBstb}. Combining these facts and using the second relation in \eq{tisigom}, we have the constraint on $\lambda$ for stable $Q$-matter solutions to exist, namely
\bea{rbeta}
\frac{|K|e^{-1}}{n-2}\exp{\bset{-\frac{n-2}{2|K|}}}<&\beta^2&<\frac{|K|e^{-1}}{n-2},\\
\label{restbeta}\Leftrightarrow \hspace*{5pt}  \frac{|K|e^{-1}}{n-2} \frac{m^{n-4}_{pl}m^2}{M^{n-2}} \exp{\bset{-\frac{n-2}{2|K|}}} <&|\lambda|^2&<\frac{|K|e^{-1}}{n-2} \frac{m^{n-4}_{pl}m^2}{M^{n-2}},
\eea
where we have used \eq{beta2} to go from \eq{rbeta} to \eq{restbeta}. Here, the lower limit of $|\lambda|^2$ corresponds to $\tiom^2_-=0$, whilst the upper limit corresponds to $\tiom^2_-=1$. The inequality in \eq{restbeta} implies that if the coupling constant $\lambda$ of the nonrenormalisable term in \eq{sugra} is too small, then it does not support the existence of $Q$-balls, whereas a large $\lambda$ coupling leads to unstable $Q$-matter. 
With the following parameter set, $m=M=1,\; |K|=0.1,\; n=6$ and the lower/upper limits of $\beta^2$ in \eq{rbeta}, \fig{fig:gravpot} shows the inverse potentials in \eq{repsugra} and their inverse effective potentials $-\Uo$ with various values of $\omega$. The lower limit, $\beta^2=\frac{|K|e^{-1}}{4}\exp{\bset{-\frac{2}{|K|}}}$, corresponds to DVPs case with $\omega_-=0$, whilst in the upper limit, $\beta^2=\frac{|K|e^{-1}}{4}$, the potentials do not have degenerate vacua with $\omega_-=1$, hence are called NDVPs. By substituting the values of $\beta^2$ into \eq{tisigom}, we obtain the values of $\sigma_+$ indicated in \fig{fig:gravpot}. Finally we can obtain the curvature, $\mu^2(\omega)\equiv\left.\frac{d^2\Uo}{d\sigma^2}\right|_{\sigma_+(\omega)}$, evaluated at $\omega_-$ , i.e.
\be{mucurv}
\mu^2 \equiv \mu^2(\omega_-) = m^2|K|(n-2)\propto |K|,
\ee
which implies that a small logarithmic correction $|K| \ll \order{1}$ in \eq{sugra} gives an ``extremely'' flat effective potential $\Uo$ compared to the quadratic term $m^2$ around $\sigma=\sigma_+$ for a given $n\sim \order{10^{0-1}}$.

\subsection{Thin-wall $Q$-ball for $\sigma_0\simeq \sigma_+,\; R_Q \gg \delta, 1/\mu,\; D\ge 2$}\label{sect:grvthn}

For the extreme limit $\omega = \omega_-$, Coleman demonstrated that the steplike ansatz \cite{Coleman:1985ki} is applicable to the case of NDVPs because the surface effects of the thin-wall $Q$-ball in this limit are not significant. There are situations though where we would like to explore the region around $\omega = \omega_-$, corresponding to $\sigma_0 \simeq \sigma_+(\omega)$, and to do this we need to include surface effects. In \cite{Tsumagari:2008bv} we explained how to do this under the assumptions: $R_Q/\delta,\; \mu R_Q \gg 1,\; \sigma(R_Q)<\sigma_-(\omega),\; \sigma_+(\omega)\simeq \sigma_+(\omega_-)\equiv \sigma_+$, and that the surface tension $\tau\simeq \int^{\sigma_+}_0 d \sigma \sqrt{2 U_{\omega_-}}$ does not depend ``sensitively'' on $\omega$. Here, $R_Q,\; \delta$ are, respectively, the $Q$-ball core size and the shell thickness. We note that in \cite{Coleman:1977py}, Coleman assumed $\Uo\simeq U_{\omega_-}$ in the shell region, and this is equivalent to saying $\sigma(R_Q)<\sigma_-(\omega)$. In what follows we will be making use of Coleman's approach. By requiring this or $\sigma(R_Q)<\sigma_-(\omega)$, we can guarantee real values of shell thickness $\delta$ and surface tension $\tau$. The assumption, in which $\tau$ does not depend on $\omega$, is related to the assumptions: $\sigma_+(\omega)\simeq \sigma_+$ and $\Uo\simeq U_{\omega_-}$ is negligible in the shell region. 

Under these assumptions and for $D\ge 2$, we now apply the previous thin-wall analysis in \cite{Tsumagari:2008bv} to the present potential \eq{sugra}. The ansatz is given by
\be{thinpro}
\sigma(r) =
    \left\{
    \begin{array}{ll}
    \sigma_+ - s(r) &\; \textrm{for}\ 0 \le r < R_Q, \\
    \bar{\sigma}(r) &\; \textrm{for}\ R_Q \le r \le R_Q + \delta, \\
    0 &\; \textrm{for}\ R_Q + \delta < r,
    \end{array}
    \right.
\ee
where $R_Q,\; \delta$, the core profile $s(r)$, and the shell profile $\bar{\sigma}(r)$ will be obtained in terms of the underlying potential by extremising $\So$ with respect to $R_Q$. Each of the profile functions satisfies
\bea{}
	s^{\p\p}+\frac{D-1}{r}s^\p -\mu s &=&0,\\
	\bar{\sigma}^{\p\p}-\left.\frac{d\Uo}{d\sigma}\right|_{\bar{\sigma}}&=&0.
\eea
By recalling \eq{infqt}, we have previously found that \cite{Tsumagari:2008bv}
\bea{RQ}
R_Q&\simeq& \bset{D-1}\frac{\tau}{\eo};\; \; \So \simeq \frac{\tau}{D} \partial V_D >0;\; \; Q\simeq \omega \sigma^2_+ V_D,\\
\label{grvthnch} \frac{E_Q}{\omega Q}&\simeq& \frac{2D-1}{2(D-1)} -\frac{\omega^2_-}{2(D-1)\omega^2},\\
\label{grvthncls} \frac{\omega}{Q}\frac{dQ}{d\omega}&\simeq& 1-\frac{2D\omega^2 }{\omega^2-\omega^2_-} <0,
\eea
where we have taken the thin-wall limit $\omega\simeq \omega_-$ in the last inequality. Notice that our analytical work cannot apply for the $1D$ thin-wall $Q$-ball, see the first expression in \eq{RQ}.

\paragraph*{\underline{\bf NDVPs:}}

This type of potential supports the existence of $Q$-matter that corresponds to the regime $\mU\gg \mS$. The $Q$-matter can be absolutely as well as classically stable for the extreme limit $\omega\simeq \omega_-$, when the coupling constant $\lambda$ for the nonrenormalisable term in \eq{sugra} satisfies \eq{restbeta}. The characteristic slope is given by the first case of \eq{virich}, and the charge and energy are linearly proportional to the volume $V_D$.

\paragraph*{\underline{\bf DVPs:}}

With the presence of degenerate minima in \eq{sugra}, in \cite{Tsumagari:2008bv} we obtained the ratio $\mU/\mS \sim 1$, which corresponds to the second case of \eq{virich}. The charge and energy are not proportional to the volume $V_D$ itself in this case; hence, we cannot see the existence of $Q$-matter in the extreme limit $\omega=\omega_-=0$. Instead we can find the proportional relation simply from \eq{leg2} and \eq{grvthnch}, namely $E_Q\propto Q^{2(D-1)/(2D-1)}$.

\vspace*{15pt}

Our main approximations are based on the assumptions $\sigma_0\simeq \sigma_+, R_Q \gg \delta,\; 1/\mu,$ and $\Uo\simeq U_{\omega_-}$ in the shell region. In what follows we will see through numerical simulations that our analytic results agree well with the corresponding numerical results even in a ``flat'' potential choice $|K|=0.1,\; m=M=1,\; n=6$, which implies that $1/\mu \sim 1.58$, see \eq{mucurv}.

\subsection{Thick-wall $Q$-balls for $\beta^2 \lesssim |K|\lesssim \order{1}$}\label{thickgrav}

In \cite{Tsumagari:2008bv} we studied thick-wall $Q$-balls in general polynomial potentials, and extracted out the explicit $\omega$ dependence from the integral in $\So$ by reparameterising terms in the Euclidean action $\So$ in terms of dimensionless quantities and by neglecting higher order terms. We then made use of the technique \eq{leg} and obtained consistent classical and absolute stability conditions, \eqs{QBstb}{QBcls}. For our present potential, \eq{repsugra}, which satisfies the condition, $\beta^2 \lesssim |K|\lesssim \order{1}$, we will be able to ignore the nonrenormalisable term  by introducing $\tisig=\sigma/M$ and $\beta^2$ in \eq{beta2}. We can then obtain the stability conditions using the same technique as before. Indeed for the limit $\omega \gtrsim \order{m}$, we will see $\tisig(r) \sim \order{\epsilon} < \order{1}$ where $\epsilon$ is a small dimensionless constant (not $\epsilon_\omega$ in \eq{infqt}), and see $\tisig_0 \equiv \tisig(0) \geq \tisig(r)$ for any $r$ because $\tisig(r)$ is a monotically decreasing function in terms of $r$. Since the leading order of the logarithmic term, $\tisig^2 \ln{\tisig}$, in \eq{repsugra} is of $\order{\epsilon^2}$ using the L'H\^{o}pital's rules, we can ignore the nonrenormalisable term in \eq{repsugra} at the beginning of our analysis. To confirm this, in Appendix \ref{appxthick} we will keep all terms in \eq{repsugra} by introducing a Gaussian ansatz and show that the results below [\eqs{grvqeq}{grvclscond}] can also be recovered under the same assumption $\beta^2 \lesssim |K|\lesssim \order{1}$. By adapting the techniques introduced in \cite{Tsumagari:2008bv}, in this subsection we will show how to obtain the thick-wall solutions without involving the Gaussian ansatz.

First of all we introduce two characteristic limits: the ``moderate limit'' $\omega\gtrsim \order{m}$ and the ``extreme'' limit $\omega \gg m$. We will see $\tisig_0\simeq \tisig_-(\omega)\to 0^+$ which leads to $\tisig_-(\omega) \ll \order{1}$ in the ``extreme limit'', and then even in the ``moderate limit'' we will see that the contributions from the nonrenormalisable term are negligible and that $\tisig_-(\omega)$ is a monotonically decreasing function in terms of $\omega$. Under the conditions $\beta^2 \lesssim |K|\lesssim \order{1}$ in \eq{repsugra}, we obtain
\bea{thckom1}
\bset{\frac{\omega}{m}}^2&=&1-2|K|\ln{\tisig_-(\omega)}+2\beta^2 \tisig^{n-2}_-(\omega) \sim 1-2|K|\log{\tisig_-(\omega)},\\
\label{thckom3} \frac{|K|m^2}{2\omega \tisig_-(\omega)}\frac{d\tisig_-(\omega)}{d\omega}&=&  \sbset{-1+2(n-2)\frac{\beta^2 \tisig^{n-2}_-(\omega)}{|K|}}^{-1}\sim -1 <0,\\
\label{thckom2} &\Leftrightarrow & \omega \gtrsim \order{m},\; \; \tisig_-(\omega) \sim \exp{\sbset{\frac{\mo^2}{2|K|m^2}}} \to 0,
\eea
where we used $\Uo(\tisig_-(\omega))=0$ to obtain \eq{thckom1}. It follows that $\tisig_-(\omega)\ll \order{1}$ for the thick-wall limit $\omega \gg m$, and we can ignore the nonrenormalisable term. Since \eq{thckom3} implies that $\frac{d\tisig_-(\omega)}{d\omega}<0$ in the limit $\tisig_-(\omega) < \order{1}$, $\tisig_-(\omega)$ is a monotonically decreasing function. Therefore, we can ignore the contributions from the nonrenormalisable term up to $\omega\gtrsim \order{m}$ which we call the ``moderate limit'' with the notion '$\sim$' as seen in the second relations of \eqs{thckom1}{thckom3}, instead of the ``extreme'' limit $\omega \gg m$ with the notion '$\to$'. Thus, we obtain the desired results of the second relation in \eq{thckom2}. From \eq{thckom1}, the logarithmic term may be of $\lesssim \order{1}$ for $|K|< \order{1},\; \beta^2\ll \order1$ in the ``moderate'' limit, which implies that the ``moderate limit'' is valid even when $\omega \sim \order{m}$. 

Let us define $\alpha(r)$ and $\tir$ through $\tisig(r)=a\alpha(r)$ and $r=b\tir$ where $a$ and $b$ will be obtained in terms of the underlying parameters. By substituting these reparamerised parameters $\alpha,\; \tir$ and neglecting the nonrenormalisable term into \eq{euc} due to 'the L'H\^{o}pital's rules', we obtain
\bea{}
\nonumber \So &\sim& \Omega_{D-1}\int d\tir \tir^{D-1}b^D \left\{ \half \bset{\frac{aM}{b}}^2\bset{\frac{d\alpha}{d\tir}}^2 \right. \\
\label{sothick} & & - \left. \half m^2 a^2 M^2 \bset{1-\bset{\frac{\omega}{m}}^2- 2|K|\ln{a}}\alpha^2 + \half m^2|K|a^2 M^2\alpha^2\ln{\alpha^2} \right\}, \\
\label{sothick2} &=& a^2 M^2 b^{D-2}\tilde{S}(\alpha),
\eea
where $\tilde{S}\bset{\alpha(\tir/b)}\equiv \Omega_{D-1} \int d\tir \tir^{D-1} \set{\half\bset{\frac{d\alpha}{d\tir}}^2 -\half \alpha^2(1-\ln{\alpha^2})}$, which is independent of $\omega$. In going from \eq{sothick} to \eq{sothick2} we have set the coefficients of the three terms in the brackets of \eq{sothick} to be unity in order to explicitly remove the $\omega$ dependence from the integral in $\So$. In other words, we have set $a=e^{-1/2}\exp\sbset{\frac{\mo^2}{2|K|m^2}} \sim e^{-1/2}\tisig_-(\omega),\; b=\frac{1}{m\sqrt{|K|}}$. Following \eq{leg}, we can differentiate \eq{sothick2} with respect to $\omega$ to obtain $Q$ and then use the Legendre transformation to obtain $E_Q$. Coupled with \eqs{QBstb}{QBcls} we obtain both the classical and absolute stability conditions. This is straightforward and yields
\bea{grvqeq}
Q&\sim& \frac{2\omega}{m^2|K|}\So,\hspace*{10pt} \frac{E_Q}{\omega Q}\sim  1+\frac{m^2|K|}{2\omega^2} \to 1,\\
\label{grvclscond}\frac{d}{d\omega}\bset{\frac{E_Q}{Q}}&\sim&1-\frac{m^2|K|}{2\omega^2}\to 1 >0,\hspace*{10pt} \frac{\omega}{Q}\frac{dQ}{d\omega}\sim 1-\frac{2\omega^2}{m^2|K|} \to - \frac{2\omega^2}{m^2|K|}<0,
\eea
where we have taken the ``extreme'' limit $\omega \gg m$ as indicated by '$\to$'. Equation (\ref{grvqeq}) implies that the characteristic slope for the thick-wall $Q$-balls are tending towards the case $\mS\ll \mU$ in \eq{virich} and \eq{grvclscond} shows that the $Q$-balls are classically stable. These results are independent of $D$. In Appendix \ref{appxthick} we will generalise the results of \eqs{grvqeq}{grvclscond} by adopting an explicit Gaussian ansatz without neglecting the nonrenormalisable term. 

Before finishing this subsection, let us comment on possibilities to have absolutely stable thick-wall $Q$-balls in the case, $|K| < \order{1},\; \beta^2 \ll \order{1}$. The results present above still hold even in the ``moderate limit'' $\omega \sim \order{m}$ for the present case. Thus, the thick-wall $Q$-balls, if they exist, can be absolutely stable when the following conditions from \eqs{QBstb}{grvqeq} are met:
\be{thckabs3}
\omega_- < m,\hspace*{10pt} \frac{\omega}{m} < \frac{1+\sqrt{1-2|K|}}{2}, \hspace*{10pt} |K| < \half,\; \beta^2\ll \order1.
\ee
It follows that for $|K| \ge 1/2$, the thick-wall $Q$-balls are always absolutely unstable. If $\omega_- \ge m$, we know $\omega > \omega_-$ in both the ``moderate'' and ``extreme'' limits, hence the thick-wall $Q$-ball is always absolutely unstable again, see \eq{QBstb}. Notice that the condition $\beta^2 \lesssim |K|$ implies $\omega_- \lesssim \order{m}$, see \eq{rbeta}, so the first condition in \eq{thckabs3}  can be satisfied. This then leaves only a small window of the parameter space for absolutely stable thick-wall $Q$-balls. In the numerical section, Sec. \ref{numerics}, we will confirm that the thick-wall $Q$-ball can be absolutely stable against decay into their own quanta by choosing suitable parameters, i.e. $\omega_-=0,\; \beta^2 \sim 1.90 \times 10^{-11}$, and $|K|=0.1$.

\subsection{Asymptotic profile for large $r$ and $\beta^2 \lesssim |K|\lesssim \order{1}$}

In order to obtain the full numerical profiles over all values of $\omega$, we should analytically determine the asymptotic profile for large $r$ in the potential \eq{sugra} which satisfies $\beta^2 \lesssim |K|\lesssim \order{1}$ as in the previous subsection. As long as the value of $r$ satisfies $r>R_\omega$ where $R_\omega$ is some large length scale and depends on $\omega$, we can assume that the friction term in \eq{QBeq} and the nonrenormalisable term in \eq{sugra} are negligible for large $r$. Hence, the $Q$-ball equation \eq{QBeq} reduces to the one-dimensional and integrable form
\be{asymgr}
\sigma^{\p\p}=\frac{d\Uo}{d\sigma},
\ee
where $\Uo\simeq \half m^2 \sigma^2\bset{1-\bset{\frac{\omega}{m}}^2 - |K|\log \bset{\frac{\sigma^2}{M^2}}}$. Equation (\ref{asymgr}) implies that the profile has a symmetry under the variation of $r$ because \eq{asymgr} does not depend on $r$ explicitly. Multiplying both sides of \eq{asymgr} by $\frac{d\sigma}{dr}$ leads to
\be{asymint}
\int^{\sigma(r)}_{\sigma(R_\omega)} \frac{d\sigma}{\sqrt{2\Uo}}=R_\omega-r,
\ee
where we have used the boundary conditions: $\sigma^\p(\infty) \to 0,\; \Uo(\sigma(\infty) \to 0)\to 0$ and $\sigma^\p(r) <0$. After some elementary algebra, the final asymptotic profile becomes 
\bea{asympro}
	\sigma(r)&=& M e^{M\mo^2/2m^2} \exp\bset{-\frac{m^2|K|M}{2} (r-r_\omega)^2},\\ 
\label{asympro2}	\frac{d}{dr}\bset{-\frac{\sigma^\p}{\sigma}}&=& m^2|K|M,
\eea
where $r_\omega\equiv R_\omega-\sqrt{\frac{\mo^2}{m^2}-\frac{2|K|}{M}\log{\bset{\frac{\sigma(R_\omega)}{M}}}}/(|K|m)$.
Equation (\ref{asympro}) is a consequence of the symmetry in \eq{asymgr} under the translation $r\to r- r_\omega$ from a Gaussian profile as seen in \eq{gauss} of Appendix \ref{exactsol}. Furthermore, \eq{asympro2} depends on the parameters $m,\; M,\; |K|$ in \eq{sugra}. We will later use the relation \eq{asympro2} as a criterion  that must be satisfied in obtaining full numerical profiles for all values of $\omega$. 

\vspace*{15pt}
We finish this section by recapping the key results we have derived for the case of the gravity-mediated potential, \eq{sugra}, in both the thin and thick-wall limits. In the thick-wall limit, we imposed the restrictions $\beta^2 \lesssim |K|\lesssim \order{1}$ on the potential to ignore the nonrenormalisable term. In both limits, we have derived the characteristic slopes in \eqs{grvthnch}{grvqeq} and the classical stability conditions in \eqs{grvthncls}{grvclscond} and shown that the $Q$ balls are classically stable in both cases. The thin-wall $Q$-balls in DVPs are always absolutely stable, and $Q$-matter in NDVPs can be absolutely stable when the coupling constant for the nonrenormalisable term satisfies \eq{restbeta}; whilst absolutely stable $Q$-balls in the thick-wall limit may exist only for \eq{thckabs3}. Finally, we obtained the general asymptotic profile, \eq{asympro}, for large $r$.

\section{Gauge-mediated potential}\label{sect:gauge}
\label{gauge-mediated}

\begin{figure}[!ht]
  \begin{center}
	\includegraphics[angle=-90, scale=0.5]{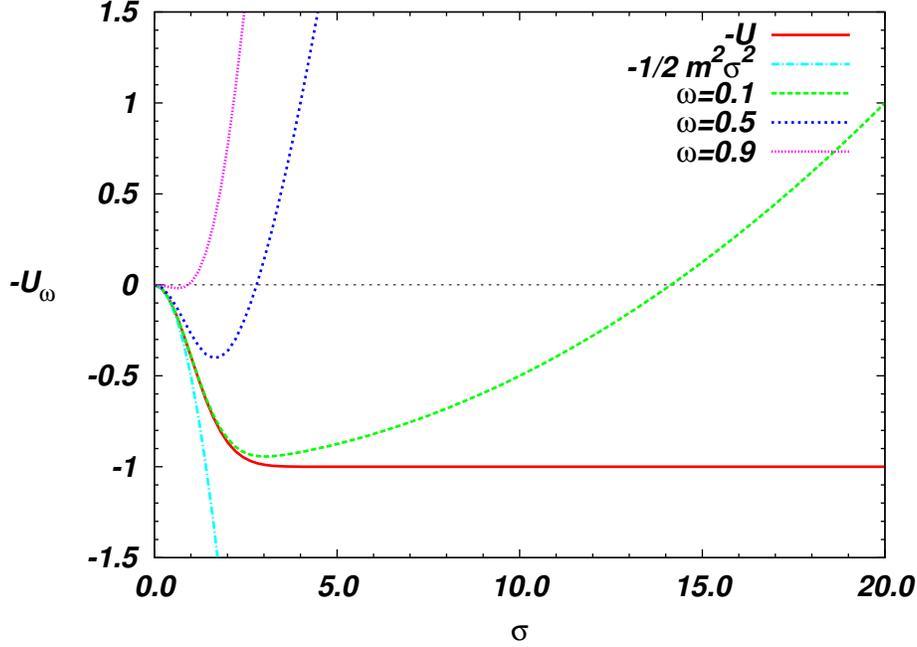}
  \end{center}
  \caption{The inverse potential $-U_{gauge}$ in \eq{apprxpot} (red-solid line) with $m=1,\; \Lambda^2=2$ which implies $U_0=1$ and the inverse effective potentials $-\Uo$ for different values of $\omega$. In order to compare between \eq{potgauge} and \eq{apprxpot}, we plot the inverse quadratic potential with the sky-blue dotted-dashed line. The two potentials are asymptotically similar, but they are different around the intermediate region of $\sigma$, where $1 \lesssim \sigma \lesssim 3$.}
  \label{fig:gaupot}
\end{figure}

The Gauge mediated scalar potential can be written in quadratic form in the low energy regime for scales up to the messenger scale $M_S$, and carries a logarithmically (extremely) flat piece in the high energy regime \cite{Dvali:1997qv, de Gouvea:1997tn}. This extreme flatness means that the thin-wall $Q$-ball we used in \eq{thinpro} cannot be applied to this situation, and so we now turn our attention to $Q$-balls in extreme flat potentials. We will generalise the results of \cite{Dvali:1997qv} to an arbitrary number of spatial dimensions and show that the known $Q$-ball profiles in \cite{Dvali:1997qv, Laine:1998rg} are naturally recovered by our more general ansatz. Moreover we will investigate both the classical and absolute stability of these $Q$-balls. The gauge-mediated potential, which we will use in this section, is approximated by \cite{MacKenzie:2001av, Asko:2002phd}
\be{potgauge}
U(\sigma)=
    \left\{
    \begin{array}{ll}
    \half m^2 \sigma^2 &\; \textrm{for}\ \sigma(r) \le \sigma(R), \\
    U_0=const. &\; \textrm{for}\ \sigma(R) < \sigma(r),
    \end{array}
    \right.
\ee
where $U_0$ and $R$ are free parameters that will be determined by imposing a condition that leads to a smooth matching of the profiles at $\sigma(R),\;  U_0 =  \half m^2 \sigma^2(R)$. Notice that $Q$-balls exist within $0 < \omega < m$ in \eq{potgauge}, and the potential does not have degenerate vacua although $\omega_-\simeq 0$. Since \eq{potgauge} is not differentiable at $\sigma(R)$, we can approximate \eq{potgauge} by
\be{apprxpot}
U_{gauge}=\half m^2 \Lambda^2 \bset{1-e^{-\sigma^2/\Lambda^2}}
\ee
which we will use in the numerical section, Sec. \ref{numerics}. Note that $\Lambda=\sigma(R)$ corresponds to the scale below which SUSY is broken, so that $U_0=\half m^2 \Lambda^2$ in \eq{potgauge}. The potential \eq{apprxpot} differs from the one used in \cite{Campanelli:2007um}, but is similar to the potential used in \cite{Gumrukcuoglu:2008gi}. \fig{fig:gaupot} shows the inverse potential \eq{apprxpot} and the inverse effective potentials for various values of $\omega$ with $m=1,\; \Lambda^2=2$, which implies $U_0=1$. The red-solid line shows the inverse potential of \eq{apprxpot} ($- U_{gauge}$), and the sky-blue dotted-dashed line corresponds to the inverse quadratic potential of \eq{potgauge}. For sufficiently large and small $\sigma$, the two potentials in \eqs{potgauge}{apprxpot} have similar behaviour, but we can see the difference in the intermediate region of $\sigma$ where $1 \lesssim \sigma \lesssim 3$. Hence, we can expect that profiles around the thick-wall limit are different between the potentials since the thick-wall profiles are constructed in the particular region, $1 \lesssim \sigma \lesssim 3$; hence it may lead to the different stationary properties and stability conditions.

Using \eq{infqt}, the $Q$-ball equation, \eq{QBeq}, in the linearised potential \eq{potgauge} becomes
\bea{qbgau1}
	\sigma^{\p\p}_{core} + \frac{D-1}{r}\sigma^{\p}_{core}+\omega^2 \sigma_{core}&=&0, \; \textrm{for}\ 0 \le r < R,\\
\label{qbgau2}	\sigma^{\p\p}_{shell} + \frac{D-1}{r}\sigma^{\p}_{shell}-\mo^2 \sigma_{shell}&=&0, \; \textrm{for}\ R \le r,
\eea
where the profiles should be imposed to satisfy the boundary conditions, $\sigma^\p<0,\; \sigma(0)\equiv \sigma_0=finite,\; \sigma(\infty)=\sigma^\p(\infty)=0,\; \sigma^\p(0)=0$. The solutions are
\be{profgauge}
    \left\{
    \begin{array}{ll}
    \sigma_{core}(r)=A\ r^{1-D/2} J_{D/2-1}(\omega r) &\; \textrm{for}\ 0 \le r < R, \\
    \sigma_{shell}(r)=B\ r^{1-D/2}K_{D/2-1}(\mo r) &\; \textrm{for}\ R \le r,
    \end{array}
    \right.
\ee
where $J$ and $K$ are Bessel and modified Bessel functions respectively, with constants $A$ and $B$.
By introducing $\sigma_0$, and expanding $J_{D/2-1}(\omega r) $ for small $\omega r$ in $\sigma_{core}(r)$, and by using the condition $U_0 =  \half m^2 \sigma^2_{shell}(R)$ we obtain
\be{au0}
A=\sigma_0 \Gamma(D/2)\bset{\frac{2}{\omega}}^{D/2-1},\hspace*{5pt} U_0=\half m^2 B^2 R^{2-D}K^2_{D/2-1}(\mo R).
\ee
Since the energy density is smooth and finite everywhere, we have to impose a smooth continuity condition to the profiles 
$\sigma_{core}(R)=\sigma_{shell}(R)$ and $\sigma^\p_{core}(R)=\sigma^\p_{shell}(R)$, which from \eq{profgauge} gives
\be{ab}
\frac{A}{B}= \frac{K_{D/2-1}(\mo R)}{J_{D/2-1}(\omega R)}=\frac{\mo K_{D/2}(\mo R)}{\omega J_{D/2} (\omega R)}.
\ee 
We will see that the particular value of $\sigma_0$ does not change important features such as the stability condition and characteristic slope of the $Q$-ball solutions. Using \eq{ab} we obtain the following important identities, which we will make use of later 
\cite{gr}:
\bea{cont1}
 \omega \frac{J_{D/2}(\omega R)}{J_{D/2-1}(\omega R)}&=&\mo \frac{K_{D/2}(\mo R)}{K_{D/2-1}(\mo R)},\\
\label{cont2} \frac{J_{D/2}(\omega R) J_{D/2-2}(\omega R)}{J^2_{D/2-1}(\omega R)}&=&-\bset{\frac{\mo}{\omega}}^2 \frac{K_{D/2}(\mo R) K_{D/2-2}(\mo R)}{K^2_{D/2-1}(\mo R)},
\eea
where we used the recursion relations $J_{\mu -1}(z)+J_{\mu +1}(z)=\frac{2\mu}{z} J_{\mu}(z),\; K_{\mu -1}(z)- K_{\mu +1}(z)=-\frac{2\mu}{z}K_{\mu}(z)$ for any real $\mu$ and $z$.
We can easily find $\So^{core}=U_0 V_D + \bset{\half \sigma_{core} (R)\sigma^\p_{core} (R)} \partial V_D$ and $\So^{shell}=-\bset{\half \sigma_{shell}(R)\sigma^\p_{shell}(R)} \partial V_D$, and then using $\So=\So^{core}+\So^{shell}$ it follows that 
\be{Soth}
\So = U_0 V_D,
\ee
where we have again used the continuity relations $\sigma_{core}(R)=\sigma_{shell}(R)$ and $\sigma^\p_{core}(R)=\sigma^\p_{shell}(R)$. To find the charge $Q$, we do not make use of the Legendre relation $Q=-\frac{d\So}{d\omega}$ in \eq{leg}, because $R$ is a function of $\omega$, and is determined by \eq{cont1}. However, we can obtain $Q$ by substituting \eq{profgauge} directly into \eq{euc}:
\be{gauq}
Q=\frac{D U_0 V_D}{\omega}  \bset{\frac{K_{D/2}(\mo R)K_{D/2-2}(\mo R)}{K^2_{D/2-1}(\mo R)}},
\ee
where we have used \eqs{au0}{ab} and \eq{cont2}, as well as the relation,\\ $\int dy\ y Z^2_\mu(y)= \sbset{\frac{y^2}{2} \bset{ Z^2_{\mu}(y)-Z_{\mu-1}(y)Z_{\mu +1}(y)}}$, \cite{Asko:2002phd, gr}. Here, $\mu$ is real, and $Z$ can be either the Bessel function $J$ or the modified Bessel function $K$, and we have used the following recursion relations to obtain the indefinite integral: $z\frac{dJ_\mu}{dz}\pm \mu J_\mu=\pm z \ J_{\mu\mp 1},\; J_{\mu-1}-J_{\mu+1}=2\frac{dJ_\mu}{dz},\; z\frac{dK_\mu}{dz}\pm \mu K_\mu=-z \ K_{\mu\mp 1},\; K_{\mu-1}+K_{\mu+1}=-2\frac{dK_\mu}{dz}$.

For future reference we obtain explicit expressions for $R$ for case with an odd number of spatial dimensions. \eq{cont1} can be solved explicitly in terms of $R$ to give
\bea{gauR1}
	      \omega R&=&  \arctan \bset{\frac{\omega}{\mo}} ,\hspace*{8pt} \textrm{for} \ D=1,\\
\label{gauR3} \omega R&=& \pi-\arctan \bset{\frac{\omega}{\mo}} , \hspace*{8pt} \textrm{for} \ D=3,
\eea
where we have used $J_{3/2}(x)=\sqrt{\frac{2}{\pi x}} \bset{\frac{\sin(x)}{x} -\cos(x)},\; J_{1/2}(x)=\sqrt{\frac{2}{\pi x}}\sin(x),\; J_{-1/2}(x)=\sqrt{\frac{2}{\pi x}}\cos(x),\; K_{3/2}(x)=\sqrt{\frac{\pi}{2x}}e^{-x}\bset{1+\frac{1}{x}},\; K_{1/2}(x)=\sqrt{\frac{\pi}{2x}}e^{-x}=K_{-1/2}(x)$. 
We will discuss the classical stability for $Q$-balls in $D=1,\; 3$ in the numerical section, in which we will show stability plots arising from \eqs{gauR1}{gauR3}.

\subsection{Thin-wall-like limit for $\mo R, \omega R \gg \order{1}$}

We now discuss both the classical and absolute stability of gauge-mediated $Q$-balls in arbitrary dimensions $D$, in the limit $\mo R,\; \omega R \gg 1$, which implies that the ``core'' size $R$ is large compared to $1/\mo,\; 1/\omega$. As we will see in the numerical section, Sec. \ref{numerics}, the limit will turn out to be equivalent to the thin-wall limit $\omega\simeq \omega_-\simeq 0$. Recall that this potential does not have degenerate vacua. Using \eqs{Soth}{gauq},
\begin{equation}
\So \simeq \frac{\omega Q}{D}\set{1+\order{(\mo R)^{-1}}},
\end{equation}
where we have used $\lim_{|z| \to \infty} K_\mu(z)\sim \sqrt{\frac{\pi}{2z}}e^{-z}\sbset{1+\frac{4\mu^2-1}{8z}+\order{z^{-2}}}$. The characteristic slope follows
\be{gauthnch}
\frac{E_Q}{\omega Q}\simeq \frac{D+1}{D}
\ee
from which we see immediately from \eq{leg2} that we recover the published results of \cite{Dvali:1997qv, MacKenzie:2001av}, namely $E\propto Q^{D/(D+1)}$. From \eqs{QBstb}{gauthnch}, the thin-wall-like $Q$-ball is absolutely stable since the present limits will cover the thin-wall limit $\omega\simeq \omega_-\simeq 0$ as we stated.

We can also obtain an explicit expression for $R(\omega)$ and $\frac{dR}{d\omega}$ in the limits $\mo R \gg 1$ and $\omega R \gg |\mu^2-\frac{1}{4}|$, where $\mu\; ( \sim \order1)$ is the argument of the Bessel function:
\bea{coregau}
\omega R &=& \bset{\frac{D+1}{4}}\pi - \arctan\bset{\frac{\omega}{\mo}},\\
\label{dRgau} \frac{dR}{d\omega}&=&-\frac{R}{\omega}\bset{1-\frac{1}{\mo R}} \simeq - \frac{R}{\omega}.
\eea
Notice that \eq{coregau} for $D=3$ reproduces the given profile in \cite{Dvali:1997qv, Laine:1998rg}, and it coincides with the exact expression derived in \eq{gauR3}. Using \eqs{gauq}{coregau} and \eq{dRgau}, we obtain
\bea{}
Q&\simeq& \frac{V_D U_0 D}{\omega},\\
\label{gaucls}\frac{\omega}{Q}\frac{dQ}{d\omega}&\simeq& -D-1 <0,
\eea
which shows that the $Q$-ball in this limit is classically stable. One can also check both $Q\simeq -\frac{d\So}{d\omega} = D U_0 V_D/\omega$ from \eq{dRgau} and  $\frac{d}{d\omega}\bset{\frac{E_Q}{Q}}\simeq\frac{D+1}{D}>0$ from \eq{gauthnch}, which are respectively consistent with \eq{gaucls} and with the result in \eq{QBcls}.

\subsection{Thick-wall limit for $D=1,\ 3,\; \dots$}

Having just discussed the thin-wall-like properties for arbitrary $D$, we turn our attention now to the  the other limit, $\omega\simeq\omega_+$. This is much more difficult to analytically explore because \eq{cont2} can only give a closed form expression for $R$ for the case where $D$ is an odd number of spatial dimensions. Therefore, we will concentrate here on the cases, e.g. $D=1,\; 3$.

\paragraph*{\underline{\bf $D=3$ case:}}

From \eq{gauR3} and recalling that in the thick-wall limit, $\mo\to0,\; \omega\simeq\omega_+=m$, we obtain $R \simeq \frac{\pi}{2\omega},\hspace{10pt} \frac{dR}{d\omega}\simeq -\frac{R}{\omega}$, and by substituting these into \eq{gauq} we find 
\bea{3dcls}
	\frac{\omega}{Q}\frac{dQ}{d\omega}&\simeq& -1 + \frac{\omega^2}{\mo^2}\to \frac{\omega^2}{\mo^2} >0,\\
\label{gauthinch} \frac{E_Q}{\omega Q}&=&1+\frac{\pi \mo}{6\omega}\to 1,
\eea
which shows that the three-dimensional thick-wall $Q$-ball is classically unstable. This fact is consistent with the relation that $\frac{d}{d\omega}\bset{\frac{E_Q}{Q}}=1-\frac{\pi\omega}{6\mo}\to-\frac{\pi\omega}{6\mo}<0$ where we have used \eq{gauthinch}. It also follows that the thick-wall $Q$-ball is not absolutely stable, and the solution will decay to free particles satisfying $E_Q\to m Q$ which is the first case of \eq{virich}.

\paragraph*{\underline{\bf $D=1$ case:}}

As in the case $D=3$, \eq{gauR1} implies $R\to 0,\; \frac{dR}{d\omega}\simeq-\frac{m^2}{\mo\omega^3}$ in the thick-wall limit. Using the above results, we obtain
\bea{1dcls}
	\frac{\omega}{Q}\frac{dQ}{d\omega} & \simeq & -1-\frac{m^2}{\omega^2}+\frac{\omega^2}{\mo^2} \to \frac{\omega^2}{\mo^2} >0,\\
\label{1dch}	\frac{E_Q}{\omega Q}&=&1+\bset{1+\frac{1}{\mo R}}^{-1}\to 1.
\eea
Note that the approximate value in \eq{1dcls} is the same as \eq{3dcls}. Then the one-dimensional thick-wall $Q$-ball is also classically unstable. This fact is again consistent with the result that $\frac{d}{d\omega}\bset{\frac{E_Q}{Q}}\simeq 1+\mo R-\frac{m^2}{\omega^2} -\frac{\omega^2R}{\mo} \to -\frac{\omega^2R}{\mo} <0$. As in the three-dimensional case, the thick-wall $Q$-ball is not absolutely stable, and the solution decays into its free particles.

\subsection{Asymptotic profile}

The asymptotic profile for the large $r$ regime in this model can be described by the contribution from the quadratic term in the potential \eq{potgauge}, from which the profile is
\be{gauasym}
\sigma(r)\sim E\sqrt{\frac{\pi}{2\mo}}r^{-\frac{D-1}{2}}e^{-\mo r} \Leftrightarrow -\frac{\sigma^\p}{\sigma}\sim \frac{D-1}{2r}+\mo
\ee
where $E$ is a constant \cite{Tsumagari:2008bv}. Note that we have used the fact that the modified Bessel function of the second kind has the relation $K_{\mu}(r)\sim \sqrt{\frac{\pi}{2r}}e^{-r}$ for large $r$ and any real $\mu$. We will use the criterion in the second expression of \eq{gauasym} in the following section.

\vspace*{15pt}

Summarising our most important results, the thin-wall-like $Q$-ball is classically stable for a general $D$, whilst it is absolutely stable as seen in \eqs{gauthnch}{gaucls}. On the other hand, for thick-wall $Q$-balls in $D=1,\ 3$, the $Q$-balls are both classically and absolutely unstable, as can be seen from \eqs{3dcls}{gauthinch} and \eqs{1dcls}{1dch}. Finally we obtained the general asymptotic profile \eq{gauasym} for large $r$.

\section{Numerical results}
\label{numerics}

In this section, we obtain exact numerical solutions for $Q$-balls for both the gravity-mediated potential in \eq{repsugra} and the gauge-mediated potential in \eq{apprxpot} with dimensionless parameters by setting $m=M=1$ and $\Lambda^2=2$. We adopt the 4th-order Runge-Kutta algorithm and usual shooting methods to solve the second order differential equations \eq{QBeq} (for full details see the  numerical techniques  developed in \cite{Tsumagari:2008bv}). The raw numerical data contains errors for large $r$, thus we introduce the previously obtained analytical asymptotic profiles to help control these uncertainties. In particular we use \eq{asympro2} for the gravity-mediated potential and  \eq{gauasym} for the gauge-mediated case. Using these techniques, the numerical profiles match smoothly and continuously onto the analytic ones. In order to check the previously obtained analytic results, we calculate $Q$-ball properties numerically over the whole parameter space $\omega$ except around the extreme thin-wall limit $\omega=\omega_-$, because it is difficult to obtain reliable numerical results in that limit.

\subsection{Gravity-mediated potential}

We shall investigate the gravity-mediated potential with two choices of $\lambda$ in \eq{sugra} for $|K|=0.1$ and $n=6$, which can be seen as the red solid lines in \fig{fig:gravpot}. The choice of the parameters, $|K|$ and $n$, are simply from phenomenological reasons. The degenerate vacua potential (DVP) on the left has $\omega_-=0$ ($\beta^2=\frac{|K|e^{-1}}{4}\exp\bset{-\frac{2}{|K|}} \sim 1.90 \times 10^{-11} \ll \order{1}$), and the nondegenerate vacua potential (NDVP) on the right has $\omega_-=1$ ($\beta^2=\frac{|K|e^{-1}}{4}\sim 9.20 \times 10^{-3} \ll \order{1}$), recalling \eq{rbeta}. \fig{fig:gravpot} also shows plots of the inverse effective potentials $-\Uo$ for various values of $\omega$. Because of numerical complications, we are unable to fully examine the properties in the extreme thin-wall limit; however, by solving close to this wall limit, our numerical results recover the expected analytical results we derived in \eqs{grvthnch}{grvthncls}. With the above choice of parameters, the curvature $\mu$ of $\Uo$ at $\sigma_+(\omega_-)\equiv \sigma_+$ in \eq{mucurv} is $\mu^2 \sim 0.4$ which implies that $1/\mu \sim 1.58$. From the first relation in \eq{tisigom}, we have found $\sigma_+\sim 1.28$ in NDVP and $\sigma_+\sim 1.91 \times 10^2$ in DVP. Since we have assumed $R_Q\gg 1/\mu,\; \sigma_0\simeq \sigma_+$ in our thin-wall analysis for the gravity-mediated potential, we see that it breaks down when the core size $R_Q$ becomes the same order as $1/\mu$ and/or $\sigma_0 \not\sim \sigma_+$. Although the full definition of the core size $R_Q$ is presented in \cite{Tsumagari:2008bv}, it is very time consuming to evaluate it properly in the simulations; hence, in this analysis we have used a more naive approach, in which we have estimated the value of $r=R_Q$ when the field profile drops quickly from its core value.  For the thick-wall limit, we required the condition $\beta^2 \lesssim |K|\lesssim \order{1}$, which is satisfied with the above chosen parameter set; hence, the analysis is valid for $\omega \gtrsim \order{1}$. Because of the choice of $|K|=0.1 < \order1$ and $\omega_-=0$ in NDVP, we will see our analysis holds even for $\omega \sim \order{1}$.

\paragraph*{\underline{\bf Hybrid profile:}}

\begin{figure}[!ht]
  \begin{center}
    \includegraphics[angle=-90, scale=0.31]{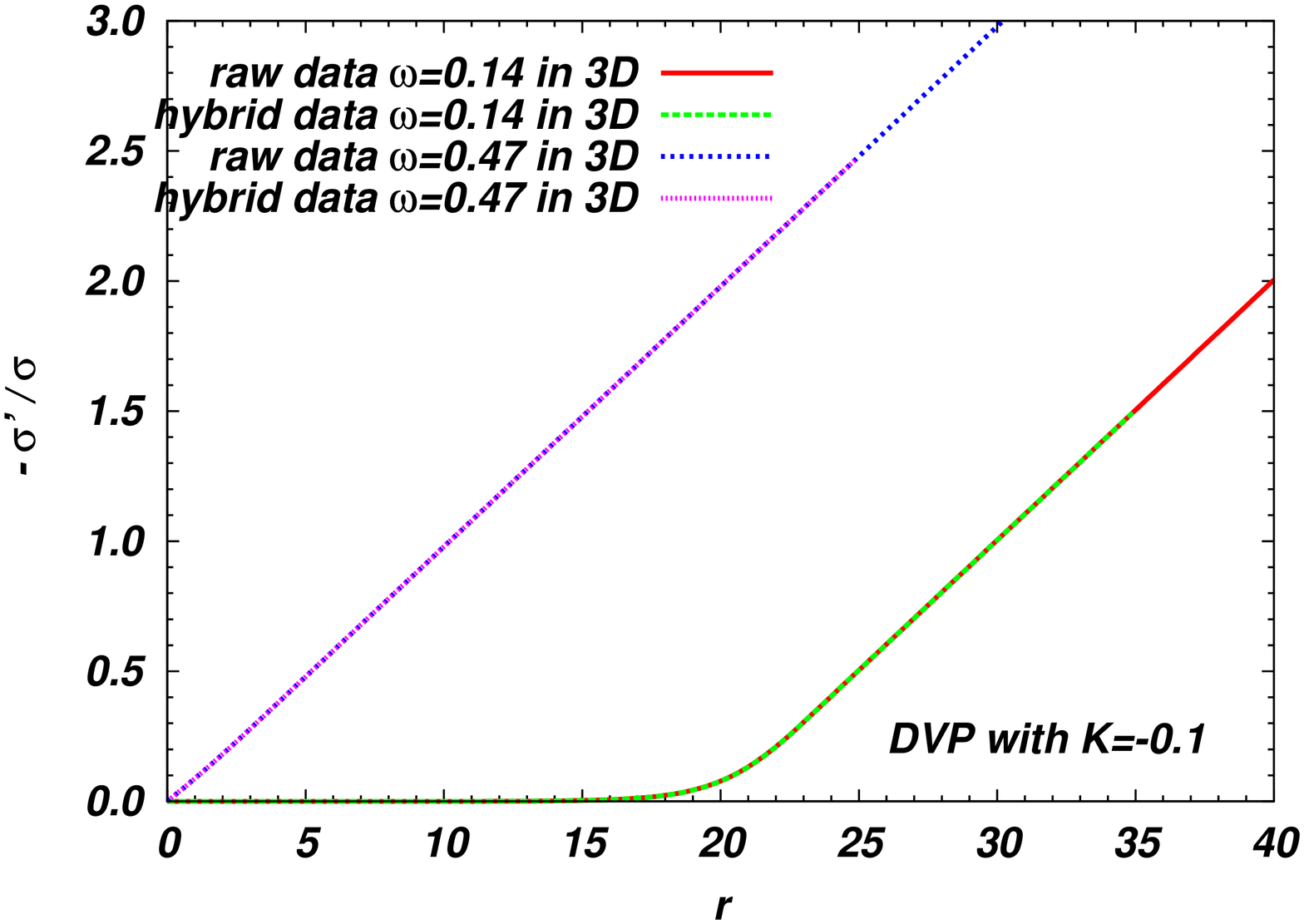} 
   \includegraphics[angle=-90, scale=0.31]{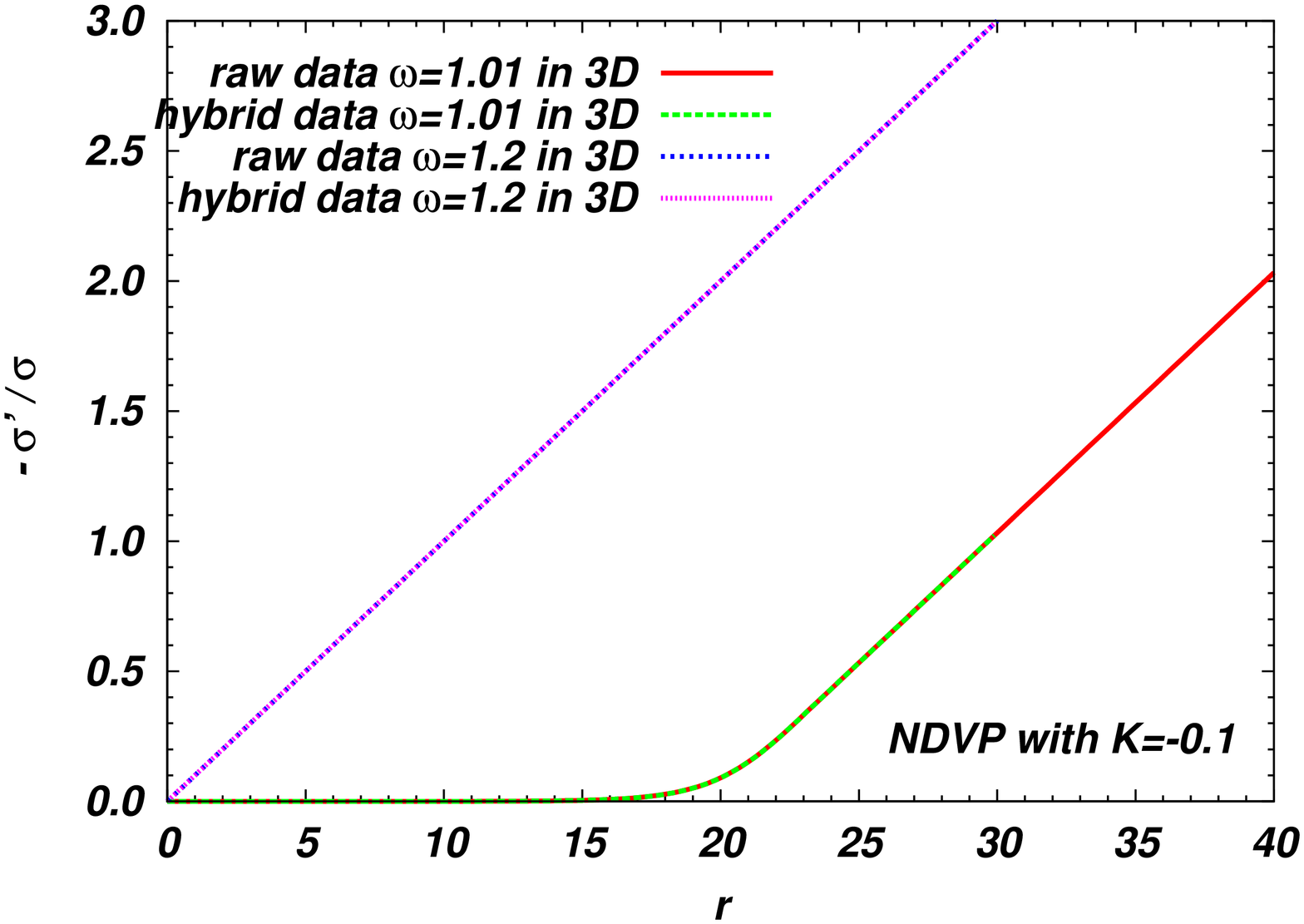} \\	
   \includegraphics[angle=-90, scale=0.31]{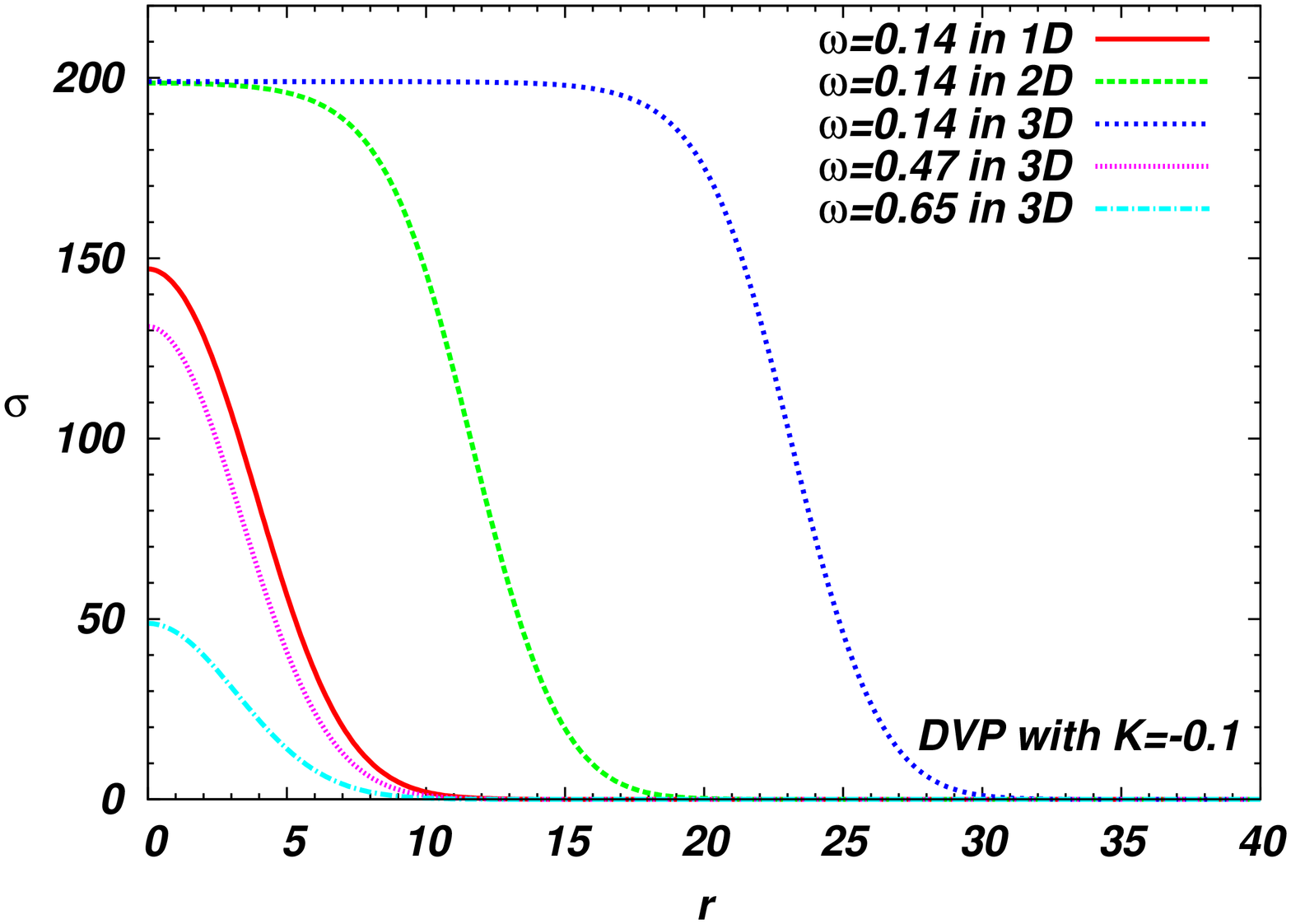} 
   \includegraphics[angle=-90, scale=0.31]{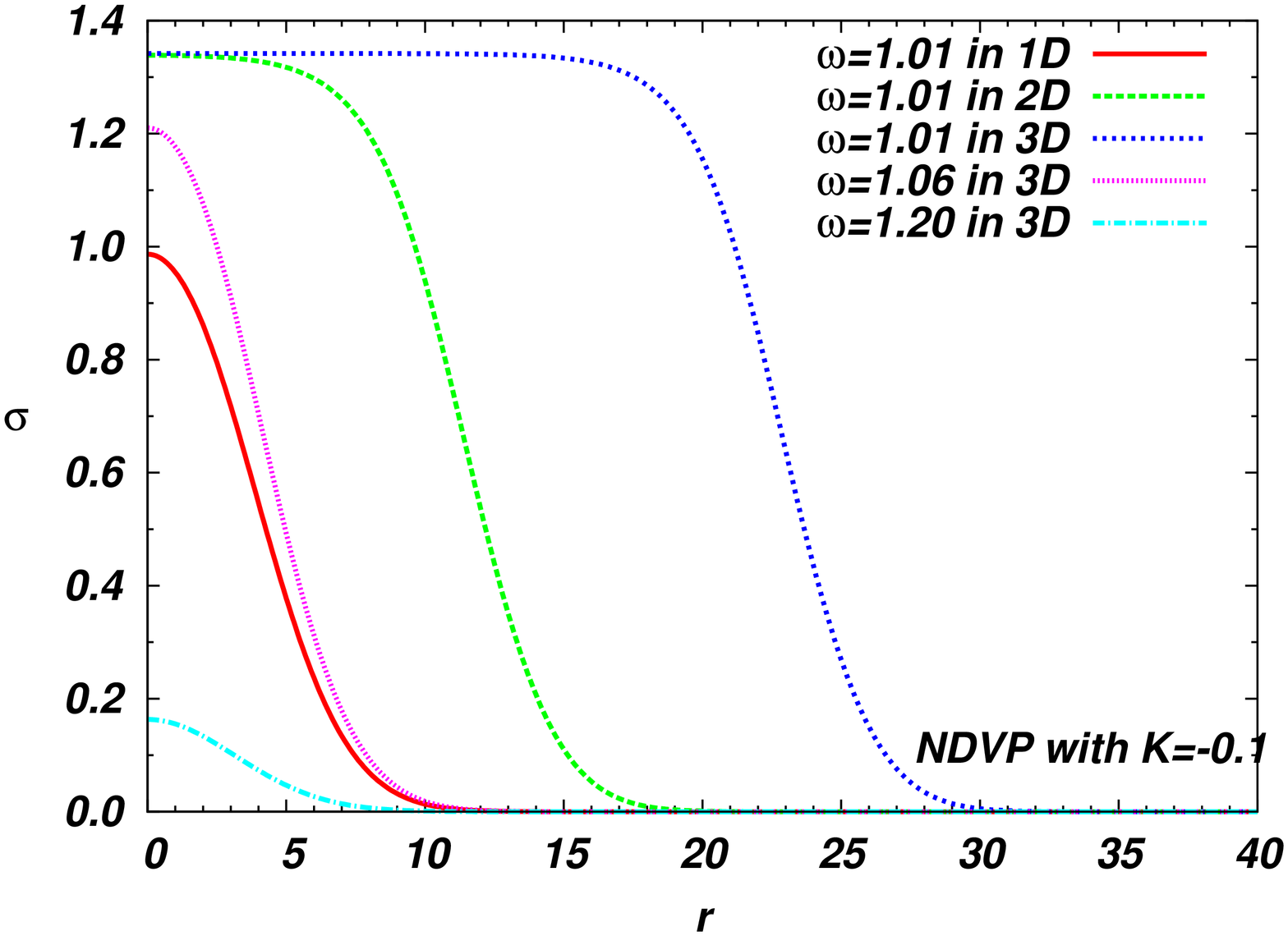} \\
   \includegraphics[angle=-90, scale=0.31]{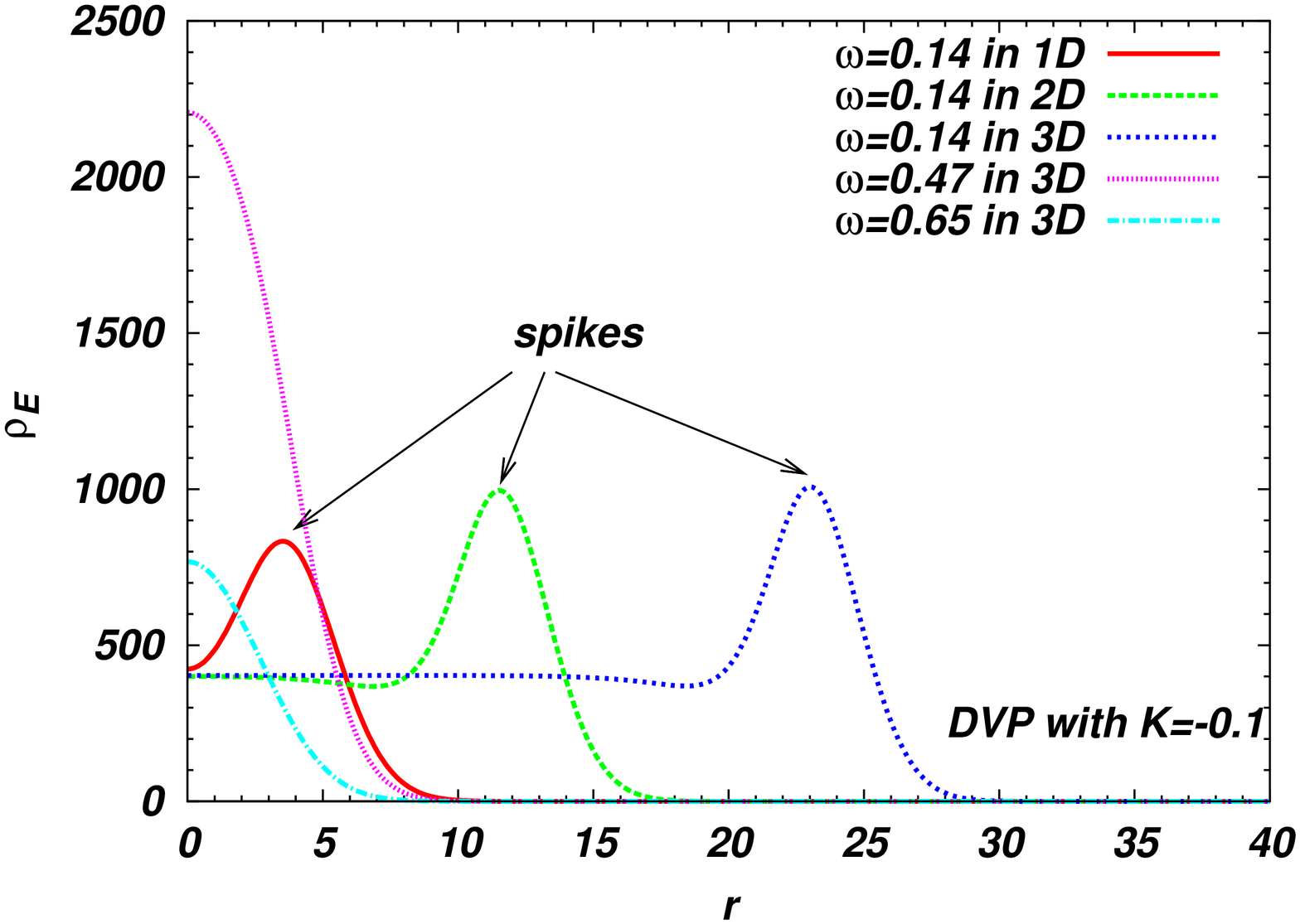}
   \includegraphics[angle=-90, scale=0.31]{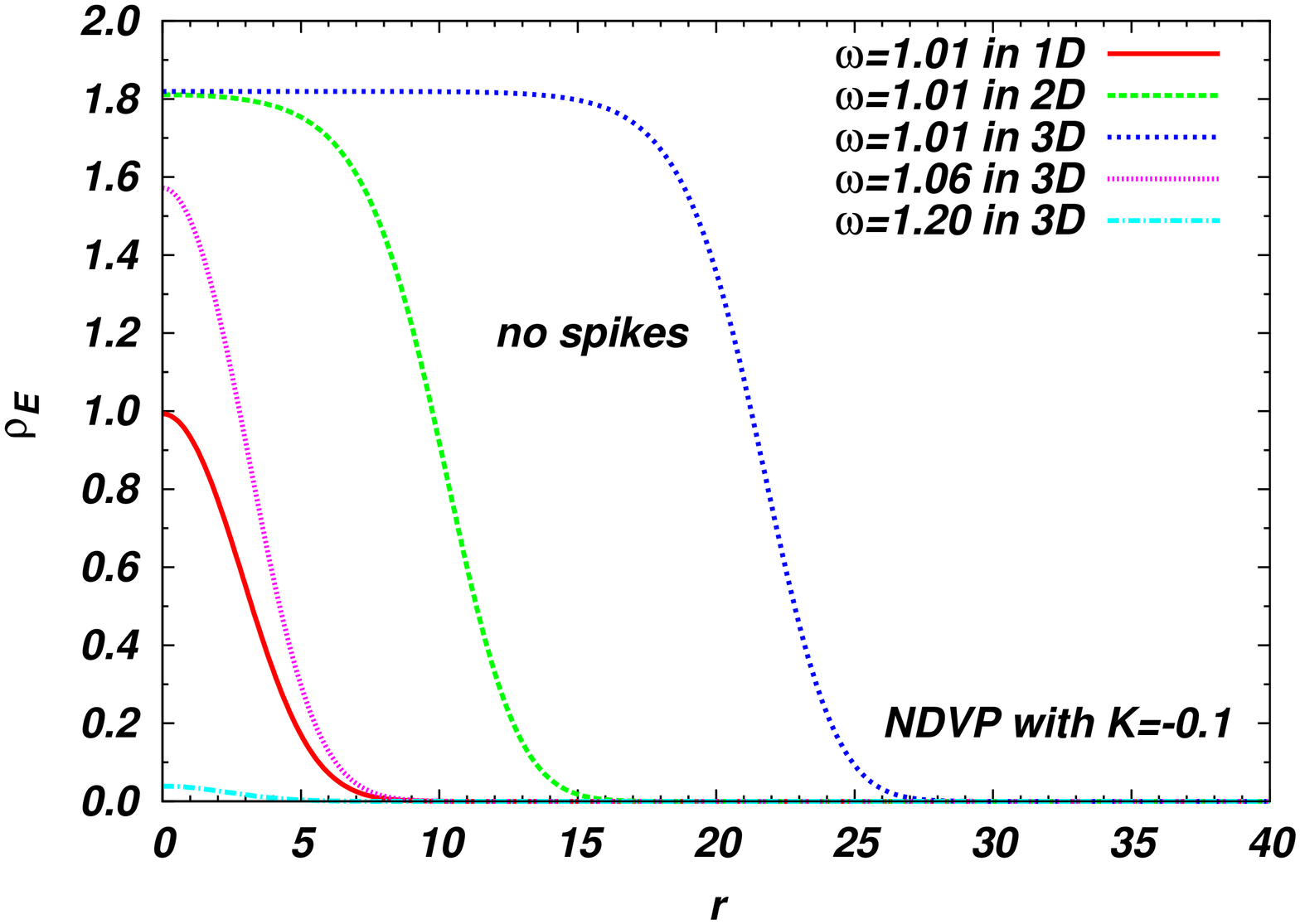}\\
  \end{center}
  \caption{The top two panels show the three-dimensional numerical slopes $-\sigma^\p/\sigma$ for two typical values of $\omega$ for both DVP (left) and NDVP (right). The raw numerical data (red-solid and blue-dotted lines) matches continuously on to the analytical asymptotic profiles for large $r$ (green-dashed and purple-dotted-dashed lines). The linear lines correspond to the Gaussian tails in \eq{asympro} where we can see the large shifts in the thin-wall limits of $\omega$. The middle and bottom panels show, respectively, the hybrid profiles \eq{hybridprof} and the energy density configurations for the various values of $\omega$ and $D$. The spikes of the energy density configurations exist in the DVP case but not in the NDVP case.}
  \label{fig:grvpro}
\end{figure}

The numerical profiles have errors for large $r$ which correspond to either undershooting or overshooting cases; thus, to minimise the errors in the region of large $r$ we replace the numerical data by the predicted asymptotic analytical profile using the criterion \eq{asympro2} to obtain the solution for the whole range of $r$. We then have the hybrid profile which can be written as
\be{hybridprof}
    \sigma(r)=
    \left\{
    \begin{array}{ll}
    \sigma_{num}(r), &\ \ \textrm{for $r<R_{num}$}, \\
    \sigma_{num}(R_{num})\exp\bset{-\frac{|K|}{2}R^2_{num}-\frac{\sigma^\p_{num}(R_{num})}{\sigma_{num}(R_{num})}R_{num} }& \\
\times \exp\bset{-\frac{|K|r^2}{2}+\bset{R_{num}|K|+ \frac{\sigma^\p_{num}(R_{num})}{\sigma_{num}(R_{num})} } r} &\ \ \textrm{for $R_{num} \le r \le R_{max}$},
    \end{array}
    \right.
\ee
where $\sigma_{num}$ is the numerical raw data, $R_{num}$ is determined by $|\bset{-\sigma^\p_{num}/\sigma_{num}}^\p -1 |_{r=R_{num}} < 0.001$, and we have set $R_{max}=60$ throughout our numerical simulations in this subsection. We have calculated the following numerical properties using the above hybrid profile, \eq{hybridprof}, for $D=1,\; 2,\; 3$:

\paragraph*{\underline{\bf Profile:}}
In the top two panels of \fig{fig:grvpro} (DVP on the left and NDVP on the right), the red-solid and blue-dotted lines show the numerical slopes $-\sigma^\p/\sigma$ for two typical values of $\omega$ in $D=3$. We smoothly continue them to the corresponding analytic profiles by the methods just  described in the numerical techniques, see green-dashed and purple-dotted-dashed lines. The linear lines correspond to the Gaussian tails in \eq{asympro} and for the cases of $\omega =0.14$ (DVP) and $\omega = 1.01$ (NDVP) corresponding to the thin-wall solution we see that it is shifted from the origin to $r \simeq 21$. The middle panels show the obtained hybrid profiles of \eq{hybridprof} for the various values of $\omega$ and $D$. The higher the spatial dimension, the larger the core size $Q$-balls can  have. The energy density configurations $\rho_E(r)$ can be seen in the bottom panels of \fig{fig:grvpro}. Outside of the cores of the DVP profiles for $\omega \sim \omega_-$, we can see the same features  as we saw in the polynomial potentials we investigated in \cite{Tsumagari:2008bv}, namely, highly concentrated energy density spikes. In NDVP, however, the spikes cannot be seen. The presence of the spike contributes to the increase in the surface energy $\mathcal{S}$, which in turn leads to the different virialisation ratio for $\mathcal{S}/\mathcal{U}$ where $\mathcal{U}$ is the potential energy, as can be seen in \eq{virich}.

\paragraph*{\underline{\bf Criterion for the existence of a thin-wall $Q$-ball:}}

\begin{figure}[!ht]  
  \begin{center}
	\includegraphics[angle=-90, scale=0.31]{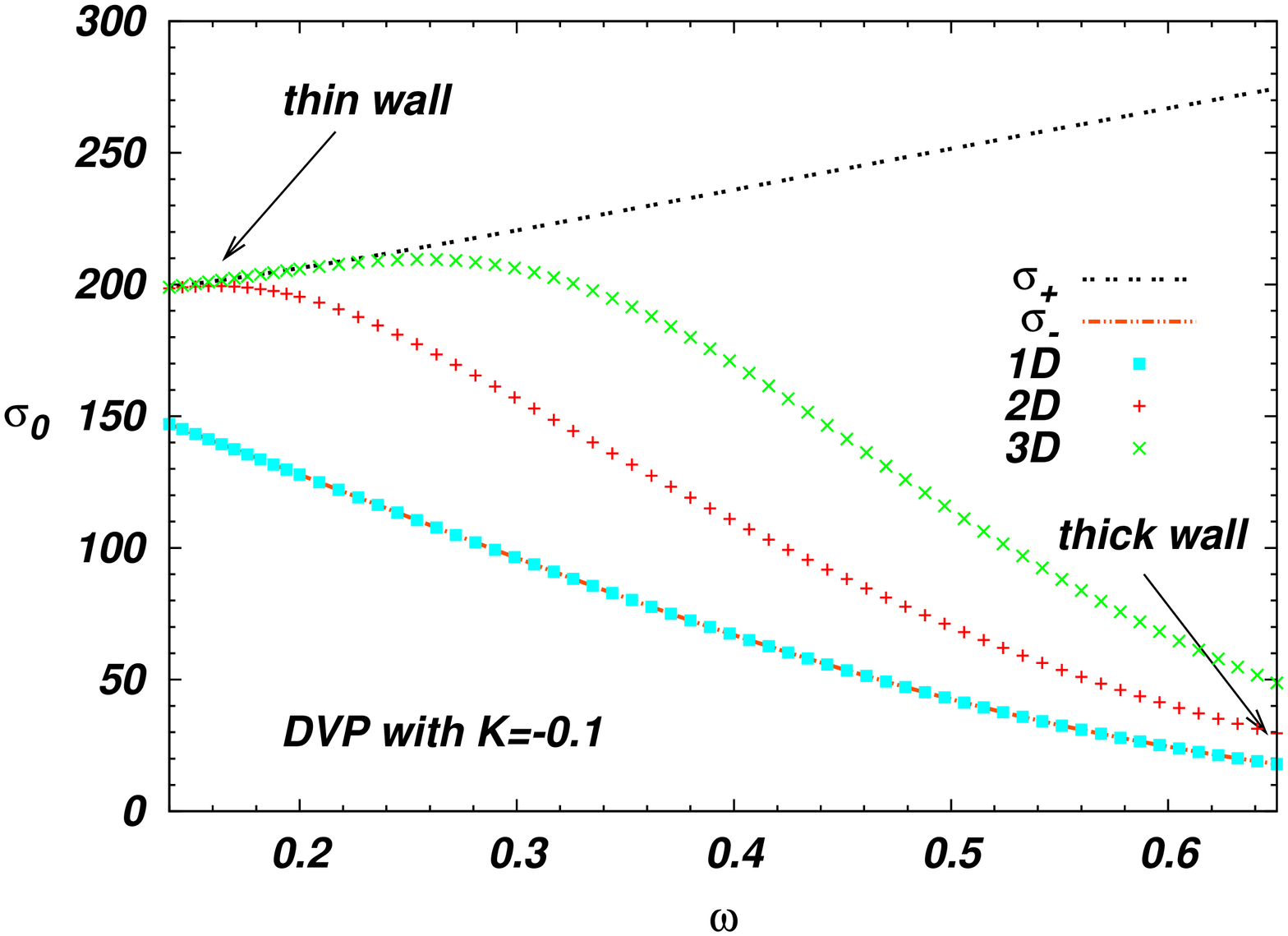}
	\includegraphics[angle=-90, scale=0.31]{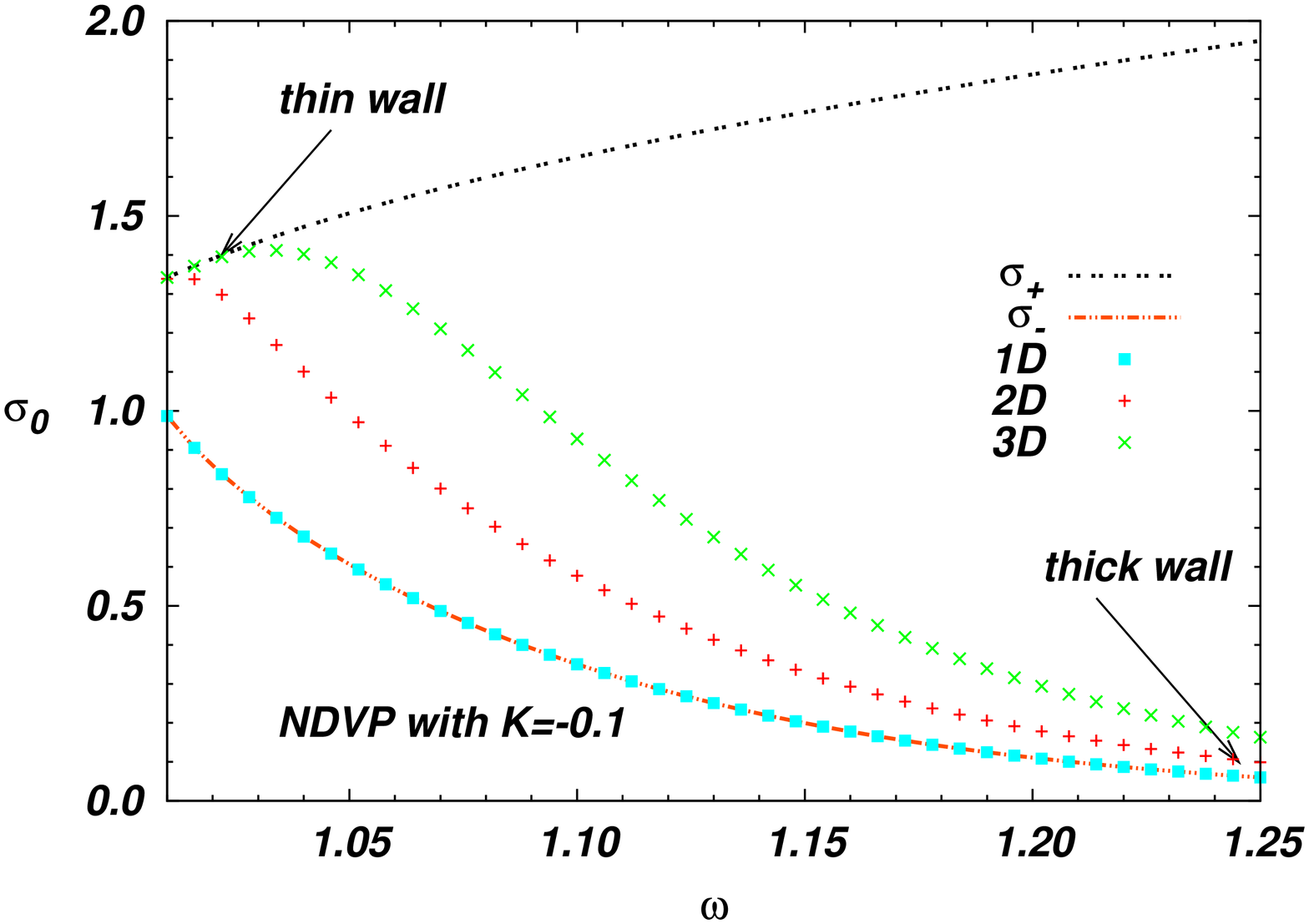}
  \end{center}
\caption{The initial value $\sigma_0(\omega)\equiv \sigma(0)$ is plotted against $\omega$. In the two panels the black-dashed and orange dotted-dashed lines show  $\sigma_\pm(\omega)$, and these lines become closer for $\omega = \omega_-$ for both types of the potentials DVP (left, $\omega_-=0$) and NDVP (right, $\omega_-=1$). Since $\sigma_0\simeq \sigma_+$ for $D=2,3$ in the region $\omega \sim \omega_-$ where $\sigma_+\sim 1.28$ in NDVP and $\sigma_+\sim 1.91\times 10^2$ in DVP, our analytical results in Sec.\ref{sect:grvthn}, are valid in this region.}
	\label{fig:grvsig}
\end{figure}

\fig{fig:grvsig} shows the numerical results for $\sigma_0(\omega)$ against $\omega$ for both types of potentials -- DVP (left) and NDVP (right). Our main analytical approximation relies on $\sigma_0(\omega) \simeq \sigma_+(\omega) \sim \sigma_+ \equiv \sigma_+(\omega_-)$, where we have found $\sigma_+\sim 1.28 \sim \order{1}$ in NDVP and $\sigma_+\sim 1.91 \times 10^2 \gg \order{1}$ in DVP. The $3D$ thin-wall $Q$-ball (green-crossed dots) appears for a wider range of $\omega$ than the $2D$ $Q$-ball (red-plus signs) in DVP as well as NDVP. For each case, the approximation can be valid, respectively, up to $\omega \sim 0.24$ or $\omega \sim 1.04$ with about $10\%$ errors for the $3D$ case. Near the thick-wall limit $\omega\simeq \omega_+$ for both potentials, we see $\sigma_0\simeq \sigma_- \to 0$. The one-dimensional values (skyblue-circled dots) always lie on $\sigma_-$. Note that in the $3D$ region $\omega \gtrsim 0.53$ for DVP, we can see $\sigma_0(\omega)\lesssim \order{10^2}$, which implies that the contribution from the nonrenormalisable term in \eqs{thckom1}{lam2},  i.e. $\beta^2 \tisig^4 \lesssim \order{10^{-3}} \ll \order{1},\; \order{|K|}$, is negligible compared to other terms in \eqs{thckom1}{lam2}. Hence, our analytic solution still holds in the limit $\omega \sim \order1$  as discussed in Sec.\ref{thickgrav}. 

\paragraph*{\underline{\bf Virialisation and characteristic slope:}}

\begin{figure}[!ht]
  \begin{center}
	\includegraphics[angle=-90, scale=0.31]{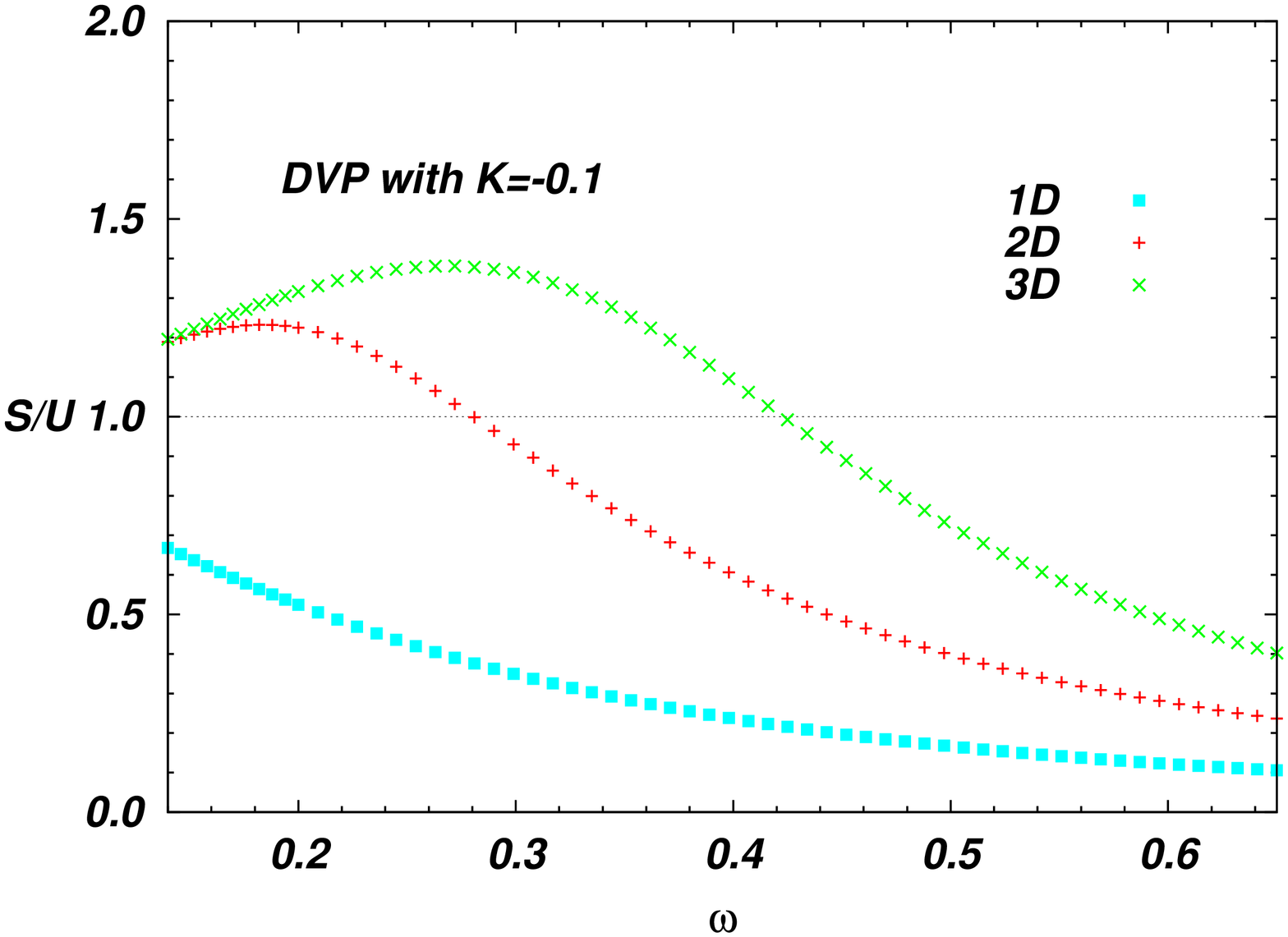}
	\includegraphics[angle=-90, scale=0.31]{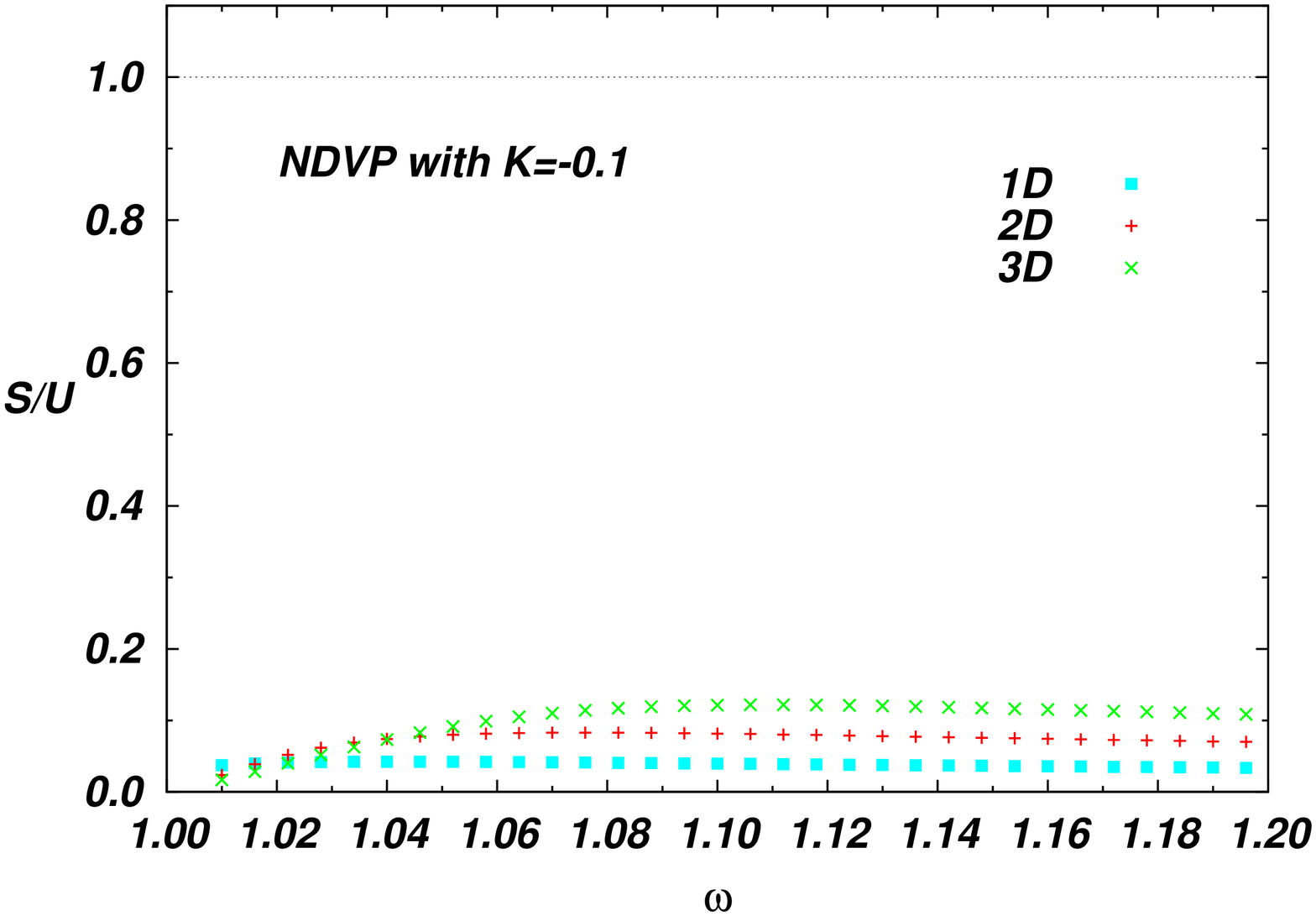}\\
	\includegraphics[angle=-90, scale=0.31]{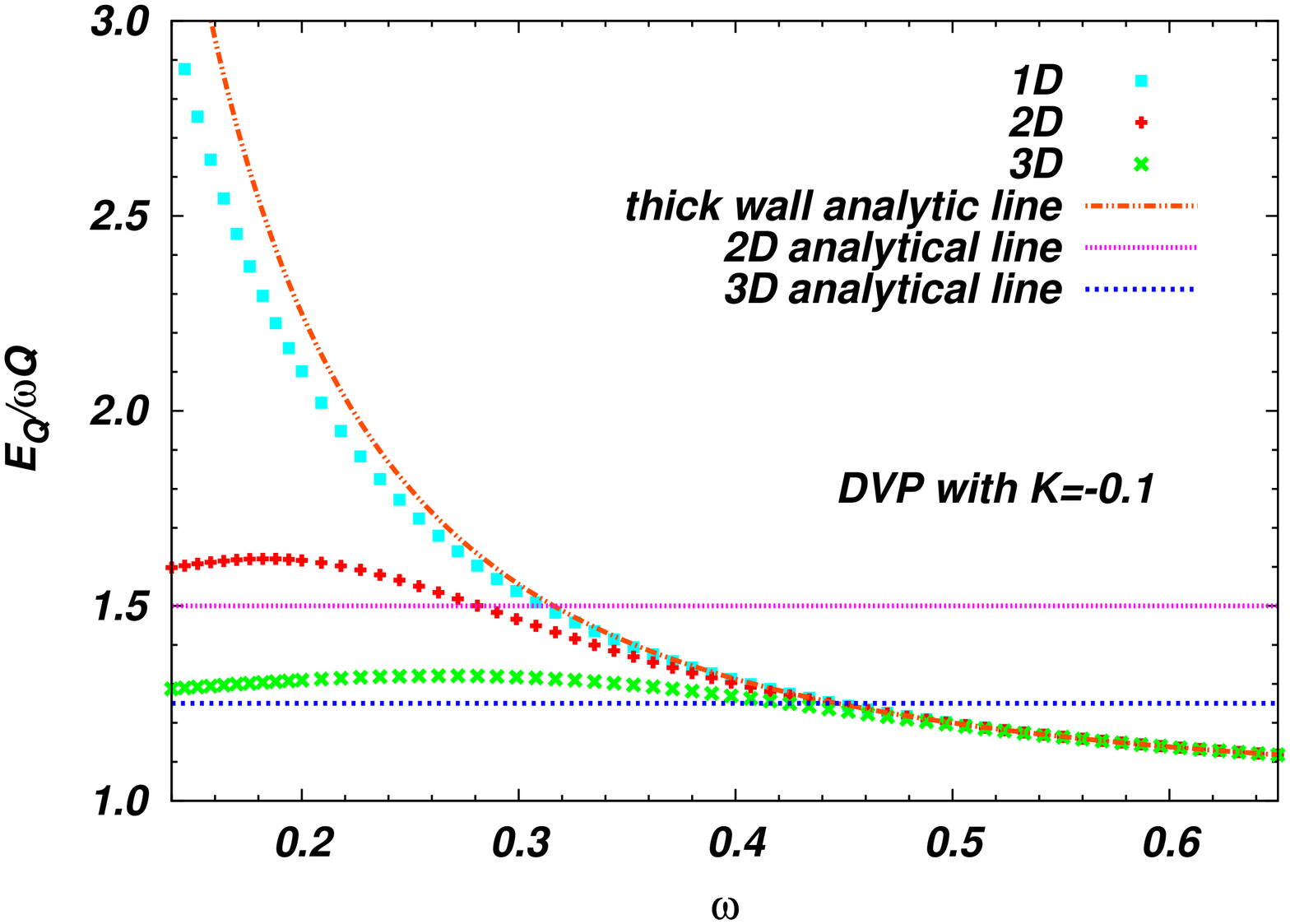}
	\includegraphics[angle=-90, scale=0.31]{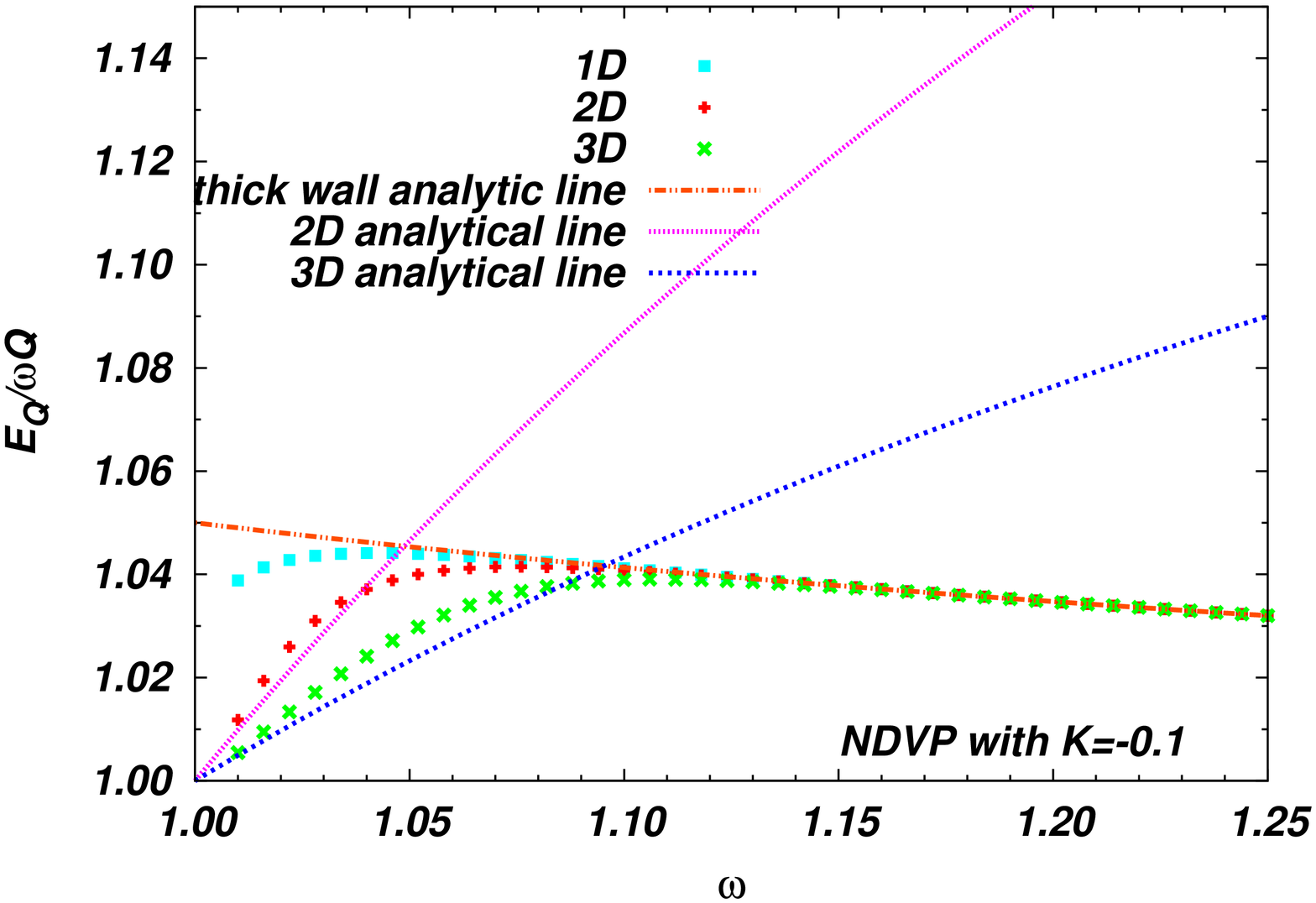}
  \end{center}
  \caption{The top panels show the ratio $\mS/\mU$ where $\mS$ and $\mU$ are the surface and potential energies, and the bottom panels show the numerically obtained  characteristic slope $E_Q/\omega Q$, in $1D$ (skyblue circled-dots), $2D$ (red plus-dots) and $3D$ (green crossed-dots). For  comparison, in the bottom panels, the thin-wall analytic lines obtained using \eq{grvthnch} are also shown (purple-dotted line for $2D$ and blue-dotted line for $3D$)  as are the thick-wall analytic lines obtained from \eqs{grvqeq}{grvch} (orange-dotted-dashed for all $D$). The analytic lines match well with the numeric data in the appropriate limits, especially for the NDVP case.}
  \label{fig:grvst}
\end{figure}

\fig{fig:grvst} shows the $Q$-ball properties plotted against the ratio of $\mS/\mU$ where $\mS$ and $\mU$ are the surface and potential energies (top panels), and the characteristic slope $E_Q/\omega Q$ (bottom panels).  For the DVP case where the thin-wall $Q$-ball satisfies $\sigma_0 \sim \sigma_+$ it appears to be heading towards  $\mS/\mU\sim 1$ as $\omega \to \omega_- = 0$ (see \eq{virich}), in all three cases. Also we predict that the thin-wall $Q$-ball in NDVP has $\mS/\mU\sim 0$ (see \eq{virich}) and that it is consistent with what can be seen in the  top right panel around $\omega=\omega_-=1$. The bottom panels  show analytically and numerically the characteristic slopes $E_Q/\omega Q$ in both the  thin and thick-wall limits. The analytic thin-wall lines (purple-dotted line for $2D$ and blue-dotted line for $3D$) based on \eq{grvthnch} are well fitted for the NDVP case with the corresponding numerical dots (red plus-dots for $2D$ and green crossed-dots for $3D$) as long as $\sigma_0\simeq \sigma_+$, see the criteria in \fig{fig:grvsig}. For the DVP case, our numerical data is seen to be heading in the right direction. The numerical solutions for both cases in the thick-wall region are well fitted by the analytic solution in general $D$ given by the orange-dotted-dashed lines, in the second relation of \eq{grvqeq} or \eq{grvch}. From the virial relation \eq{viri} for $D=1$, we can only predict the extreme values of the $1D$ characteristic slope, $\gamma$, in either the DVP or NDVP case once we know what $\mS/\mU$ is. To obtain that we rely on the numerical simulations and from the top two panels in  \fig{fig:grvst}, we see that for the DVP case with $D=1$,  $\mS/\mU$ appears to be heading towards unity, implying $\gamma \gg 1$ in \eq{viri}, whereas for the NDVP case $\mS/\mU \ll 1$, implying $\gamma \to 1$ in \eq{viri}. Comparing these with the bottom two panels we see the behaviour for $\gamma$ appears to follow these predictions.  

\paragraph*{\underline{\bf $Q$-ball stability:}}

\begin{figure}[!ht]
  \begin{center}
	\includegraphics[angle=-90, scale=0.31]{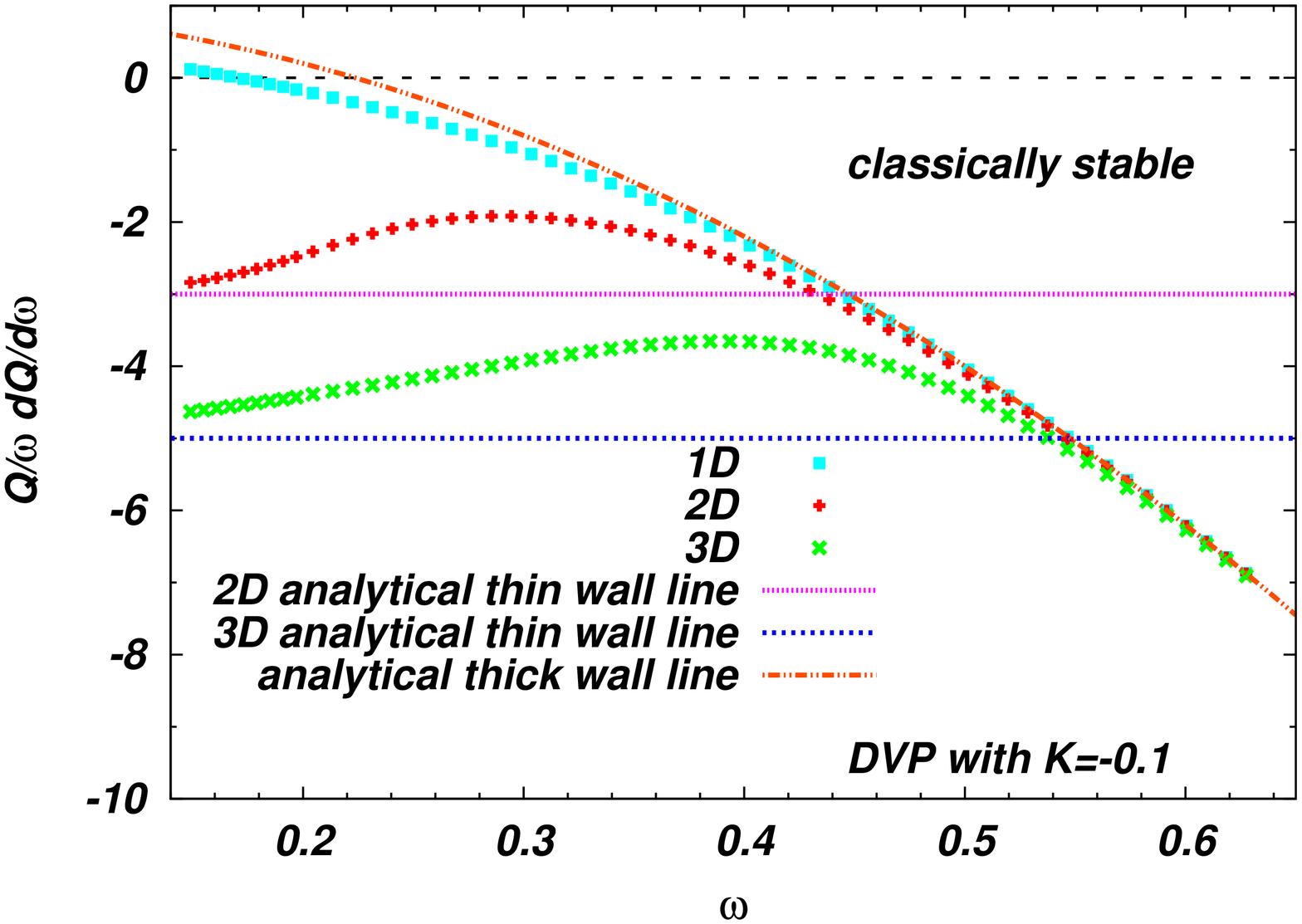} 
	\includegraphics[angle=-90, scale=0.31]{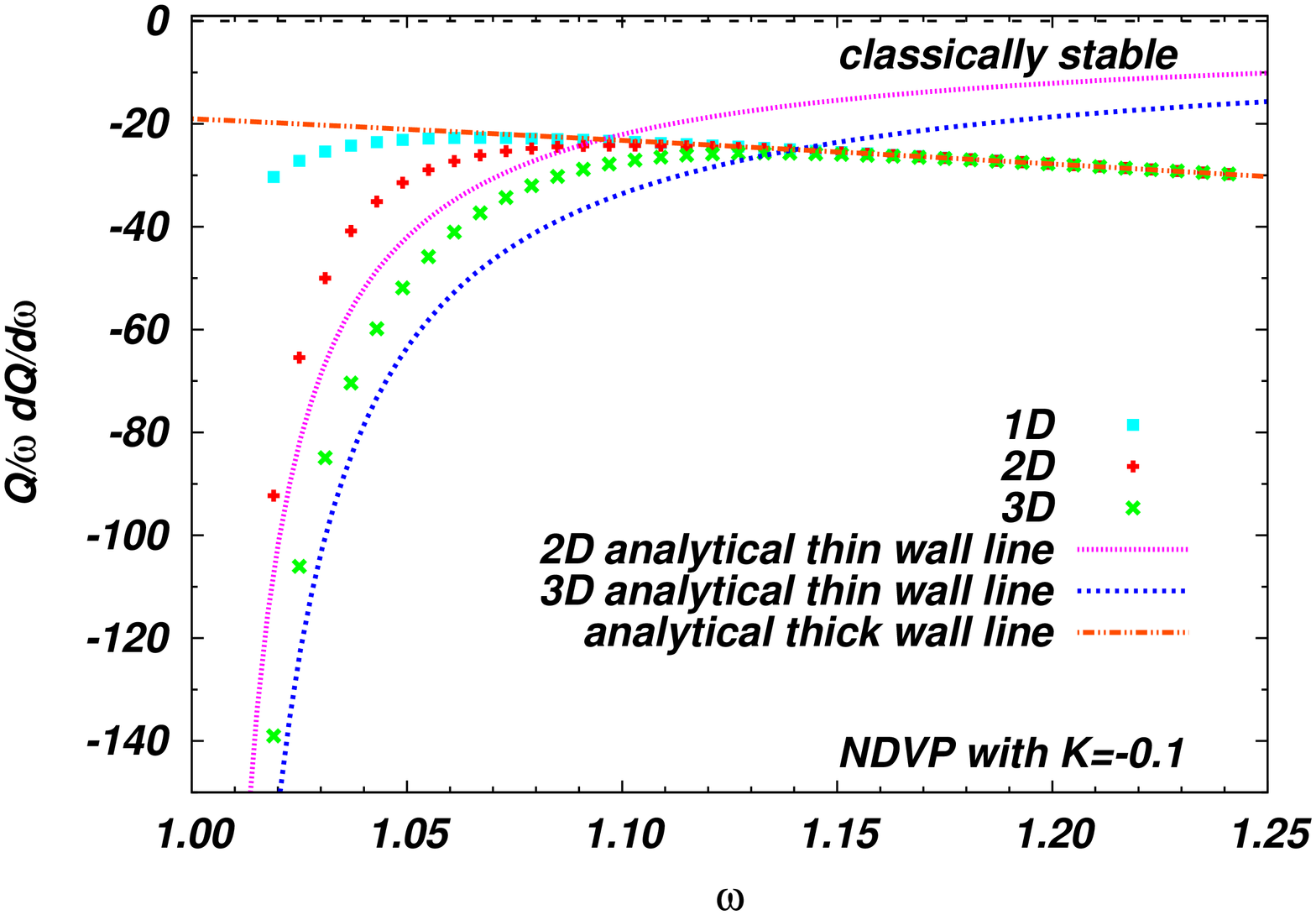} \\
	\includegraphics[angle=-90, scale=0.31]{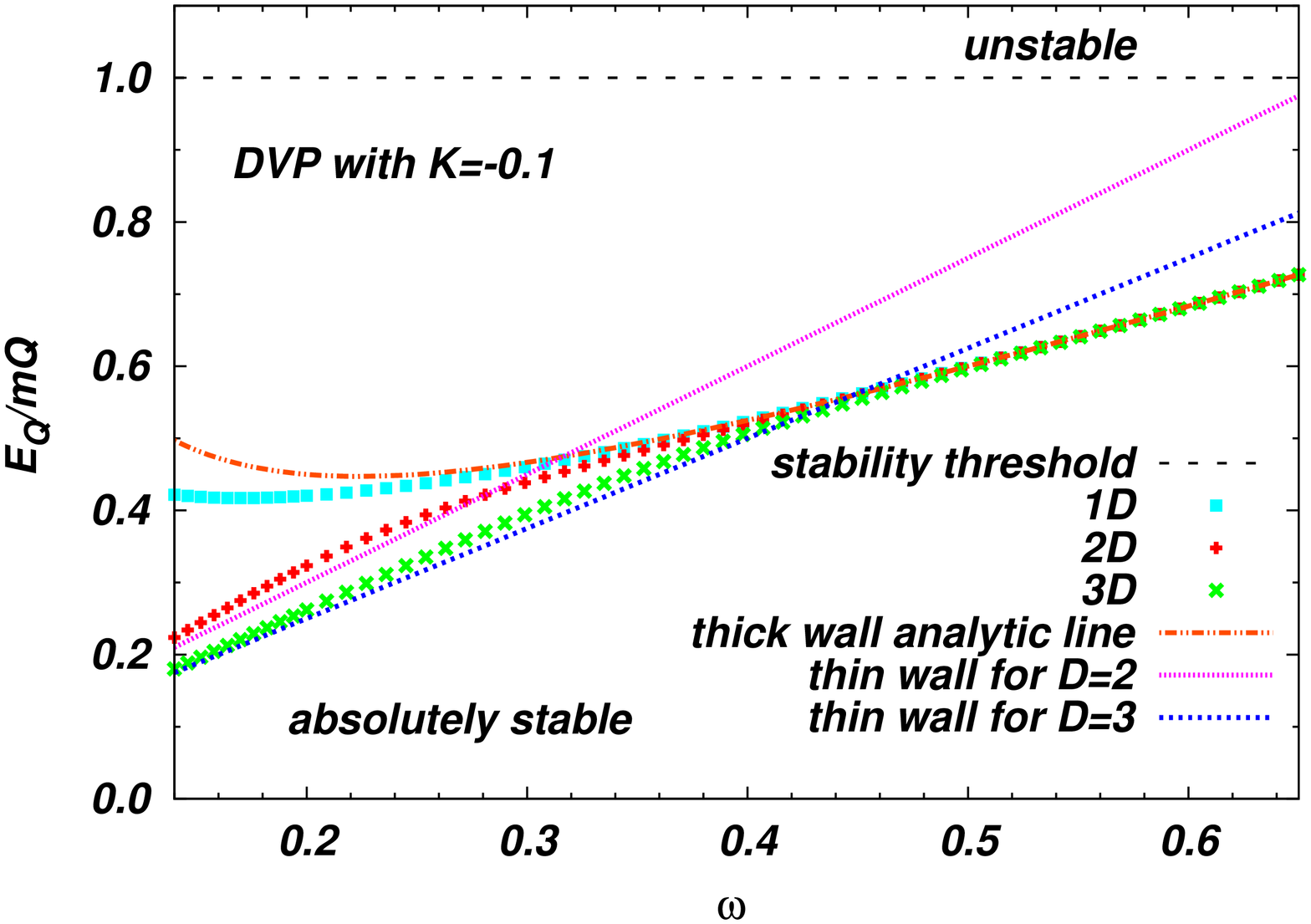} 
	\includegraphics[angle=-90, scale=0.31]{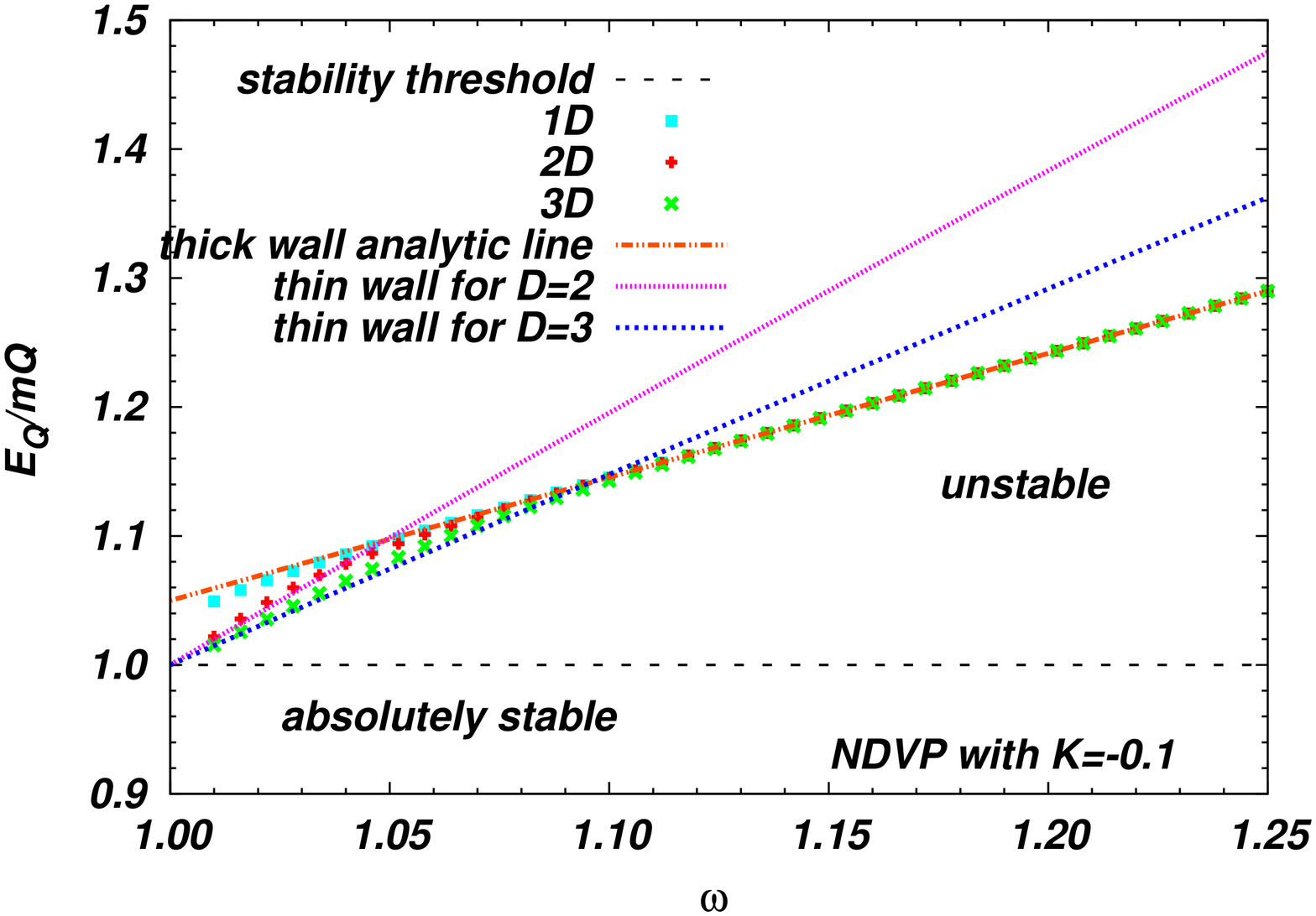}
  \end{center}
   \caption{Classical stability for the top panels and absolute stability for the bottom panels for both DVP (left) and NDVP (right). The black-dashed lines indicate the stability thresholds for both classical and absolute stability in all panels. $Q$-balls found below the lines are stable either (both) classically or (and) absolutely. In the top panels, the analytical lines using \eqs{grvthncls}{grvclscond} agree well quantitatively with the corresponding numerical data for thick-wall regimes, but not well in the thin-wall regimes. However the numerical plots look qualitatively similar to the analytical lines in the thin-wall limit as seen in the polynomial models \cite{Tsumagari:2008bv}. In addition, the analytical lines for $E_Q/mQ$ using \eqs{grvthnch}{grvch} match the numerical lines for both the thin and thick-wall limits. }
  \label{fig:grvabs}
\end{figure}

\fig{fig:grvabs} shows plots for both the classical (top panels) and absolute stability (bottom panels) with the stability threshold lines (black-dashed) for the cases of DVP (left) and NDVP (right). Let us consider the classical stability case first. For the thin-wall regime in DVP, notice that the numerical data of $\frac{\omega}{Q}\frac{dQ}{d\omega}$ (red-dot-circles for $2D$ and green-dot-crosses for $3D$) are heading towards the analytic lines of \eq{grvthncls}. For the thick-wall case, on the other hand, the analytical lines of \eq{grvclscond} (orange-dotted-dashed) fit excellently with the numerical data in all dimensions, because \eq{grvclscond} is independent of $D$. Furthermore, the $Q$-ball is classically stable over all values  of $\omega$ except for the $1D$ thin-wall case where our analytical work cannot be applied. We saw this feature of unstable $1D$ thin-wall $Q$-balls for the case of polynomial models in \cite{Tsumagari:2008bv}. For the absolute stability in the bottom panels, the analytical lines using \eq{grvthnch} and \eqs{grvqeq}{grvch} are matched with the numerical dots for both the thin and thick-wall limits. Here, we note how well the three-dimensional $Q$-ball (and also the higher dimensional ones as predicted in \cite{Tsumagari:2008bv}) can be described simply by our thin and thick-wall $Q$-balls. As our parameter set satisfies \eq{thckabs3}, we can see that absolutely stable $Q$-balls exist in DVP near the thick-wall limit. Because of the choice of $\omega_-=1$, the $Q$-ball in the NDVP case, however, is always absolutely unstable and most of the features are similar in terms of $D$. The analytical lines (top-right panel) in NDVP  agree with the corresponding numerical data qualitatively better than the lines for DVP. 

To sum up our discussion of the Gravity mediated model, our analytical estimates of the characteristic slope and other properties of the $Q$-balls are well checked against the corresponding numerical results, even though we set a ``flatter'' potential with $|K|=0.1 < \order{1}$.

\subsection{Gauge-mediated potential}

This subsection presents numerical results showing the properties of gauge-mediated $Q$-balls with $m=1,\; \Lambda^2=2$ in \eq{apprxpot}. Although we have obtained analytical results for the potential, \eq{potgauge}, the potential is neither analytic nor smooth for all $\sigma$. Therefore, we shall use the approximate potential, \eq{apprxpot}, see \fig{fig:gaupot} and we expect that \eq{apprxpot} is a suitable approximation especially for the thin-wall limit $\omega$ and large $D$. We will also see and explain the expected discrepancies that exist between the numerical and analytic results.

\paragraph*{\underline{\bf Hybrid profile:}}
As we saw in earlier examples the numerical profiles we have obtained have errors for large $r$, which correspond to either undershooting or overshooting; thus, we replace the numerical data in that regime by the exact asymptotic analytic solutions we obtained using the second relation of \eq{gauasym} to smoothly continue the numerical solutions to the corresponding analytical ones. The hybrid profile in this model is
\be{hybridgau}
    \sigma(r)=
    \left\{
    \begin{array}{ll}
    \sigma_{num}(r)&\ \ \textrm{for $r<R_{num}$}, \\
    \sigma_{num}(R_{num})\bset{\frac{R_{num}}{r}}^{(D-1)/2} e^{-\mo(r-R_{num})} &\ \ \textrm{for $R_{num} \le r \le R_{max}$},
    \end{array}
    \right.
\ee
where $\sigma_{num}$ is the numerical raw data, $R_{num}$ is determined by $|\frac{D-1}{2r}+\mo+\bset{\sigma^\p_{num}/\sigma_{num}}| < 0.001$, and we have again set $R_{max}=60$. We have calculated the following numerical properties using the above hybrid profile, \eq{hybridgau}, up to $D=3$.

\paragraph*{\underline{\bf Profile and energy density configuration:}}

\begin{figure}[!ht]
  \begin{center}
    \includegraphics[angle=-90, scale=0.50]{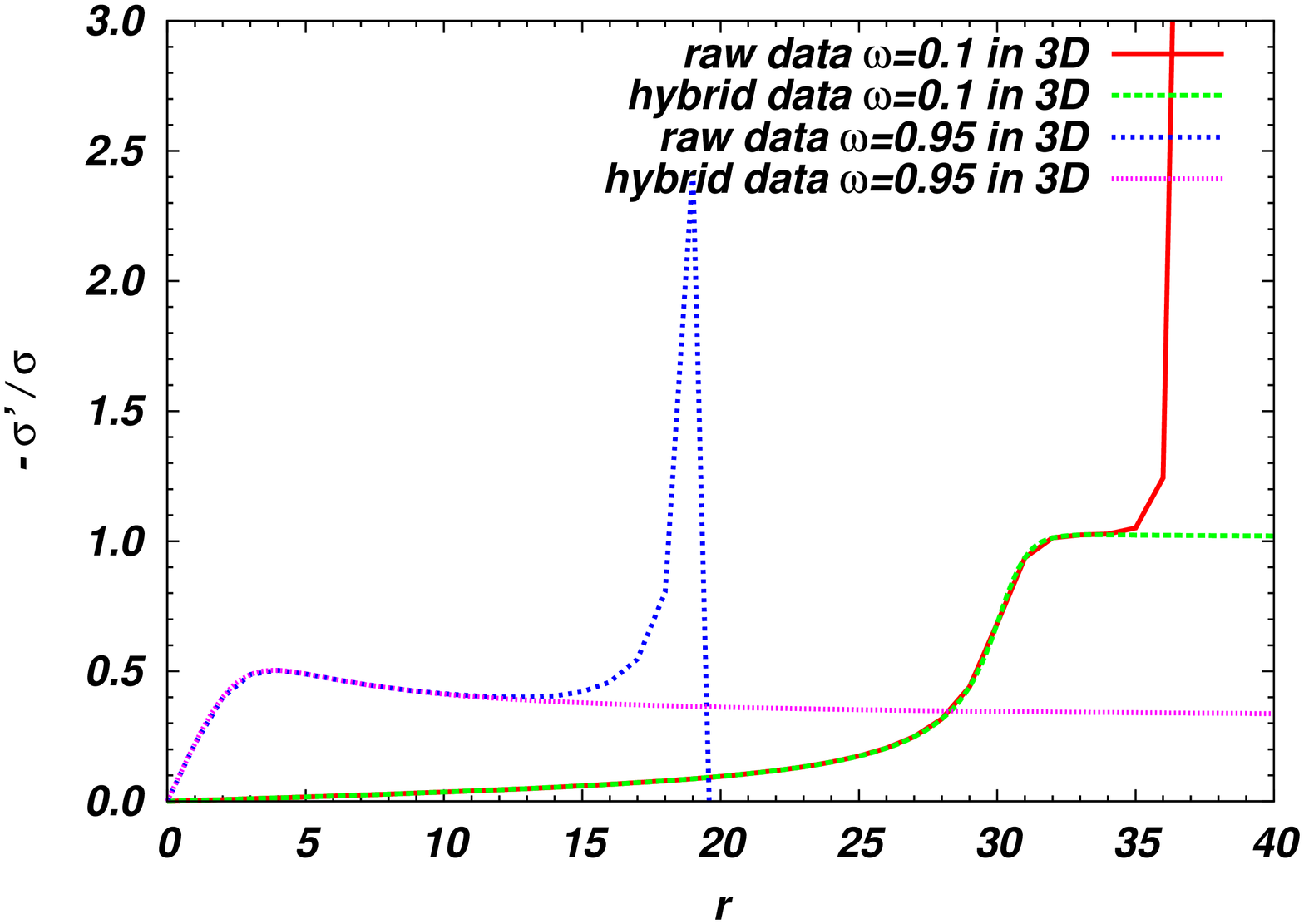}\\
   \includegraphics[angle=-90, scale=0.31]{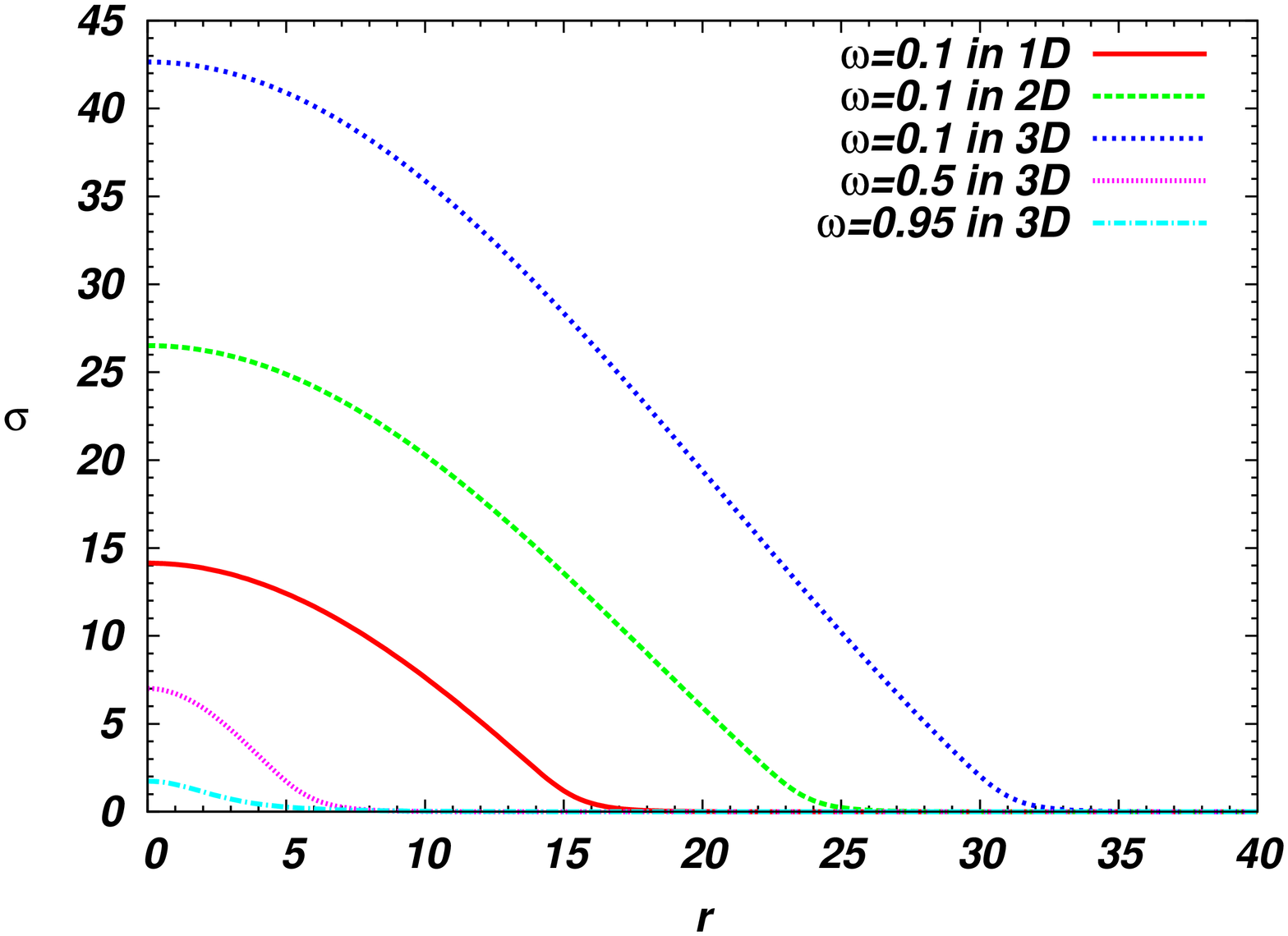} 
   \includegraphics[angle=-90, scale=0.31]{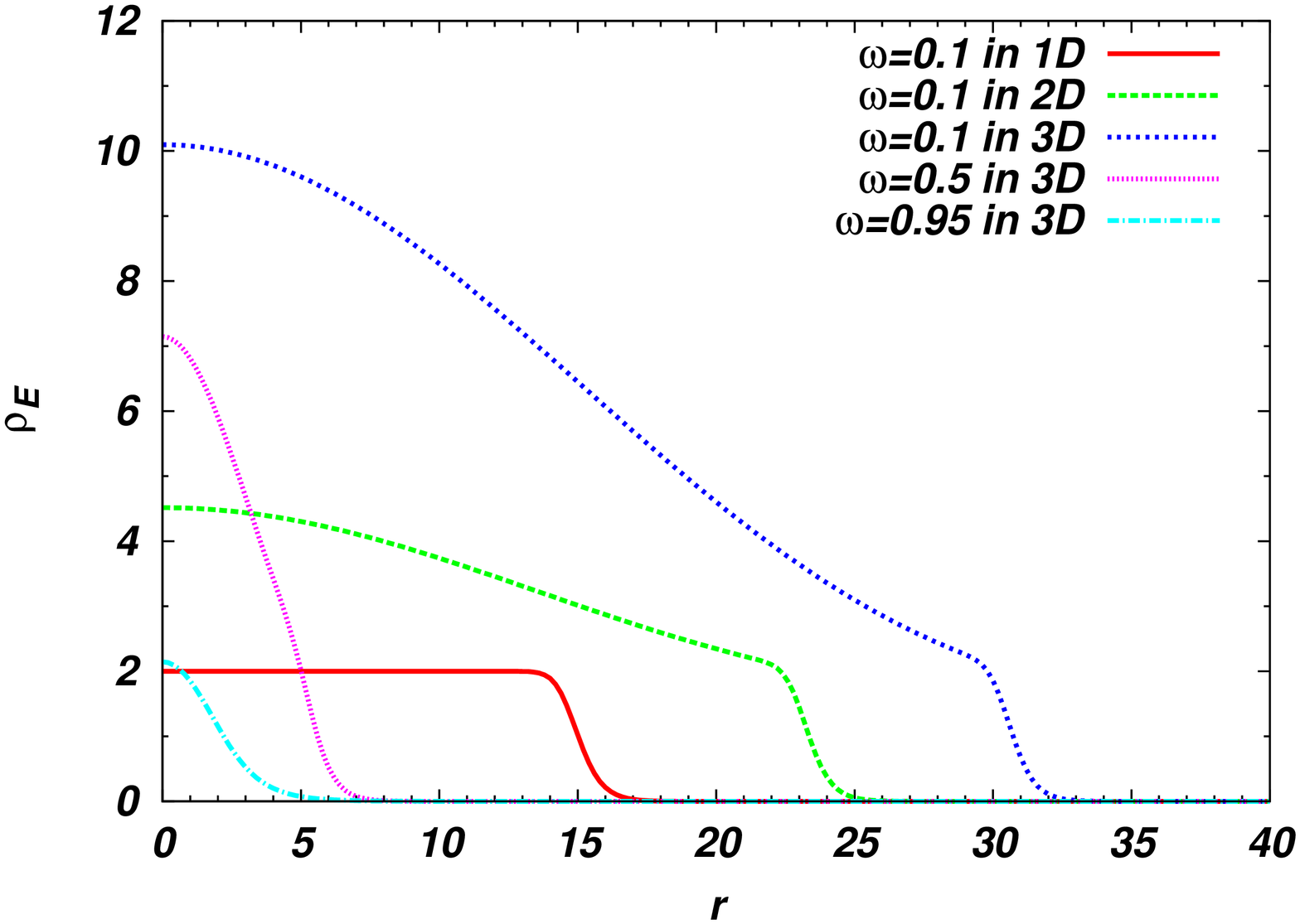}
  \end{center}
  \caption{The top panel shows the three-dimensional numerical slopes $-\sigma^\p/\sigma$ for two values of $\omega$. The raw numerical data (red-solid and blue-dotted lines) matches smoothly to the corresponding analytical asymptotic profiles for large $r$ (green-dotted and purple-dashed lines). Both the left- and right-bottom panels show, respectively, the hybrid profiles \eq{hybridgau} and the energy density configurations for the various values of $\omega$ and $D$. The spikes of energy density configurations do not exist even in the thin-wall limits. }
  \label{fig:gaupro}
\end{figure}

\fig{fig:gaupro} shows the three-dimensional numerical slopes $-\sigma^\p/\sigma$ for two values of $\omega$ (top), hybrid profiles (left-bottom) as in \eq{hybridgau}, and the configurations for energy density (right-bottom). In the top panel, the raw numerical data (red-solid and blue-dotted lines) is matched smoothly onto the continuous asymptotic profiles \eq{hybridgau} for large $r$ (green-dotted and purple-dashed lines). By fixing the numerical raw data using the technique \eq{hybridgau}, we show the profiles for various values of $\omega$ and $D$, see the left-bottom panel. Also the peaks of the energy density cannot be observed in the whole range of $\omega$, see the right-bottom panel.

\paragraph*{\underline{\bf Characteristic slope:}}

\begin{figure}[!ht]
  \begin{center}
	\includegraphics[angle=-90, scale=0.5]{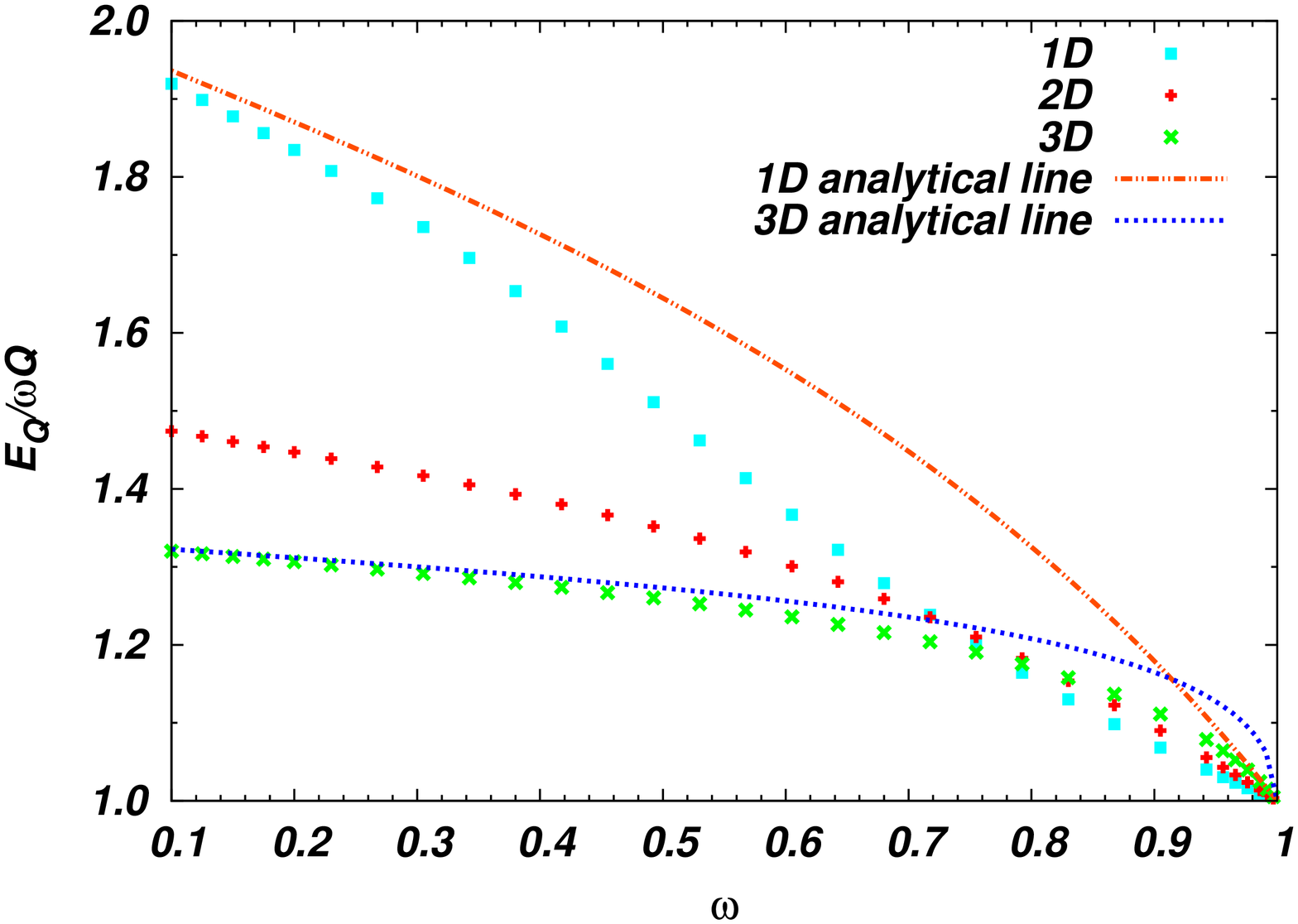}
  \end{center}
  \caption{The numeric characteristic slopes $E_Q/\omega Q$ and the analytic lines (orange-dashed line for $1D$ and blue-dotted line for $3D$) which are calculated using \eqs{gauR1}{gauR3} in the whole range of $\omega$. The $3D$ analytic line agrees with the numeric data well except near the thick-wall limit. Similarly the $1D$ analytic line agree well only in the extreme thin-wall limit.}
  \label{fig:gaueq}
\end{figure}

In \fig{fig:gaueq}, we plot both the numeric and analytic characteristic slopes $E_Q/\omega Q$ (orange-dashed line for $1D$ and blue-dotted line for $3D$). By substituting \eqs{gauR1}{gauR3} into \eqs{gauq}{euc}, we have obtained the analytic slopes covering the whole range of $\omega$. The $3D$ analytic line agrees well with the numerical data  except near the thick-wall limit. Similarly the $1D$ analytic line agrees well only in the thin-wall limit. The origin of the discrepancies in the analytic versus numerical fits are the differences between the potentials themselves [\eq{potgauge} in the analytical section (Sec. \ref{sect:gauge}) and \eq{apprxpot}]. These differences are largest between $1 \lesssim \sigma \lesssim 3$ which in turn affects the region around $0.9\lesssim \omega < 1.0$, see \fig{fig:gaupot} and \fig{fig:gaueq}. 

\paragraph*{\underline{\bf $Q$-ball stability;}}

\begin{figure}[!ht]
  \begin{center}
	\includegraphics[angle=-90, scale=0.31]{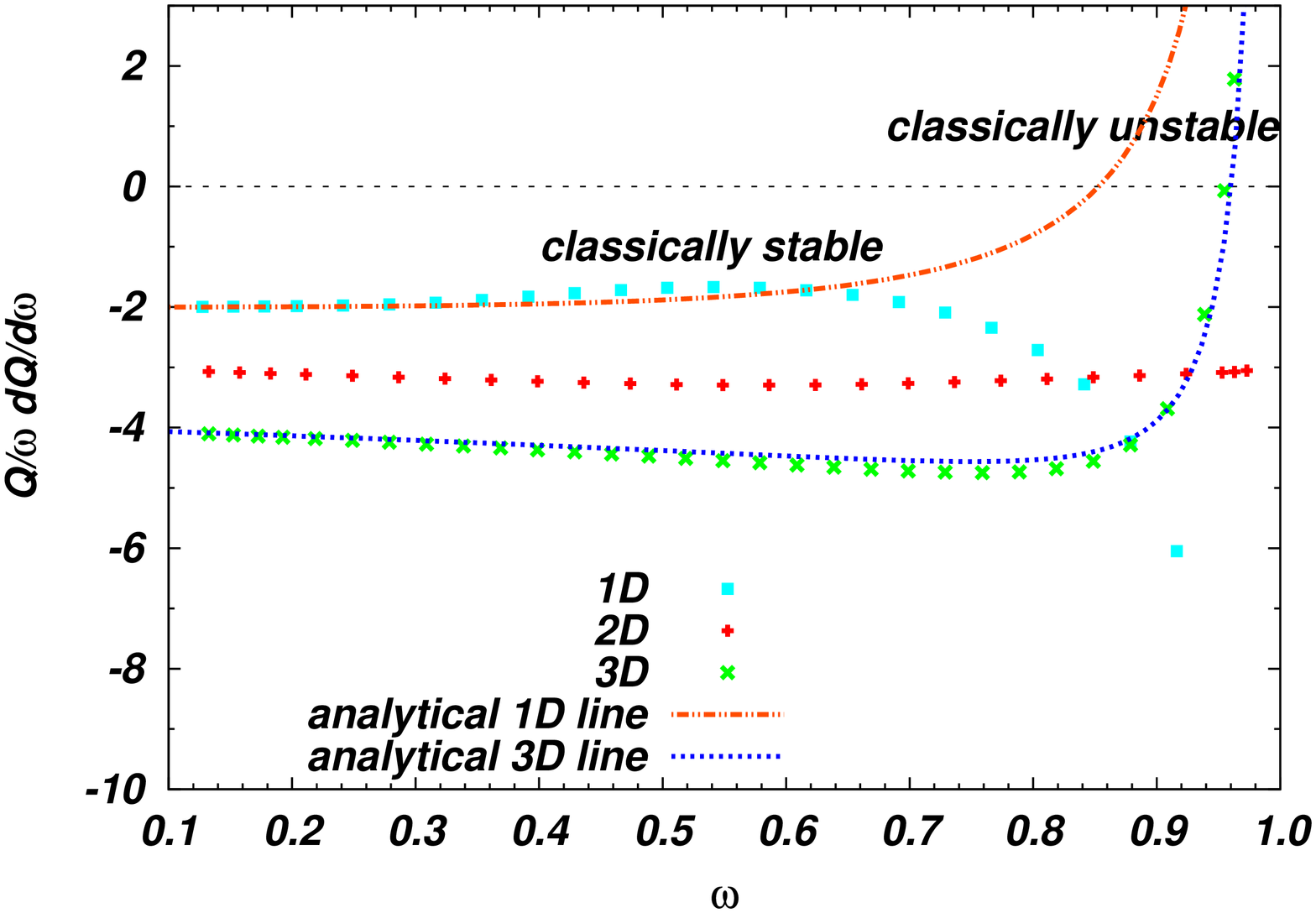} 
	\includegraphics[angle=-90, scale=0.31]{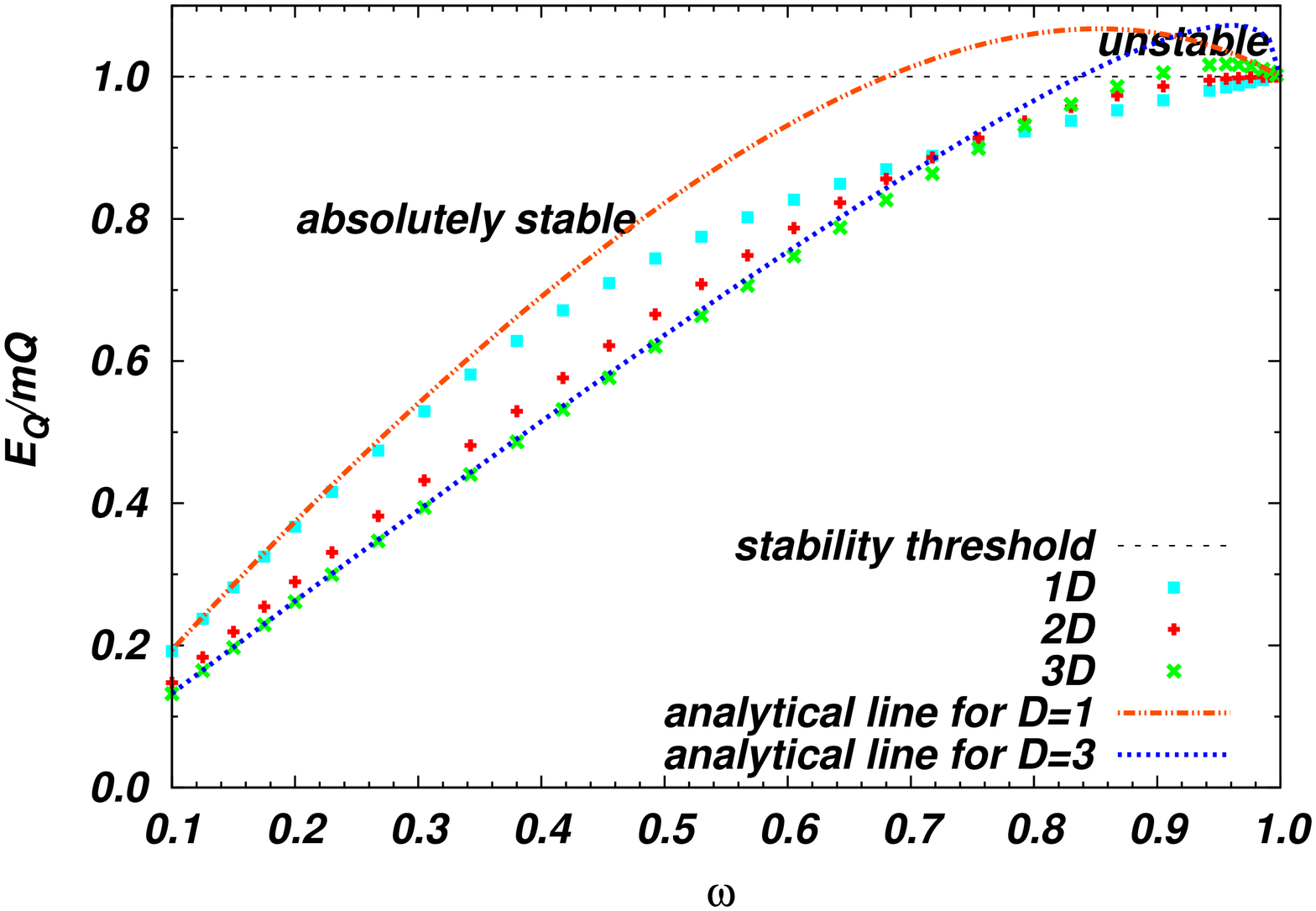} 
  \end{center}
   \caption{The stability of $Q$-balls -- Classical (left panel) and absolute (right panel). The black-dashed lines in the two panels indicate the stability thresholds for both classical and absolute stability where $Q$-balls under the lines are classically/absolutely stable. The analytic lines for $D=1,\; 3$ are calculated by substituting \eqs{gauR1}{gauR3} into \eq{gauq} and differentiating it with respect to $\omega$.}
  \label{fig:gauabs}
\end{figure}

\fig{fig:gauabs} illustrates the stability of $Q$-balls: classical stability in the left panel and absolute stability in the right panel. The black-dashed lines in both panels indicate their respective stability thresholds where $Q$-balls under the lines are stable. We calculate the analytic lines for $D=1,\; 3$ by substituting \eqs{gauR1}{gauR3} into \eq{gauq} and differentiating it with respect to $\omega$. The $3D$ numerical data can be matched with the analytic lines in both the thin and thick-wall limits. As in \eq{3dcls}, the three-dimensional $Q$-ball in the thick-wall limit is classically unstable. The numerical thick-wall $Q$-ball in $1D$ is classically stable which differs from the prediction in \eq{1dcls}. In the right panel, the analytic line for $D=3$ agrees with the numerical data except in the thick-wall limit where the analytical lines for both $1D$ and $3D$ do not match the corresponding numerical data. Furthermore, the thick-wall $Q$-ball in $1D$ is absolutely unstable as predicted analytically in \eq{1dch}, but this fact cannot be observed numerically. The reasons for this discrepancy are as before a problem with our choice of potentials. We can see that the thin-wall $Q$-balls for any $D$ are both classically and absolutely stable.

To recap, our numerical results in the Gauge mediated case are generally well fitted by our analytical estimations. Observed discrepancies between the analytical predictions and numerical data arise from the artifact of our approximated smooth potential \eq{apprxpot} for the generalised Gauge mediated potential \eq{potgauge}. We have confirmed that the thin-wall $Q$-balls for any $D$ are both absolutely and classically stable.

\section{Conclusion and discussion}
\label{conc}

We have explored stationary properties of $Q$-balls in two kinds of flat potentials, which are the gravity-mediated potential, \eq{sugra}, and the generalised gauge-mediated potential, \eq{potgauge}. Generally, the gauge-mediated potential is extremely flat compared to the gravity one; therefore, we cannot apply our thin-wall ansatz \eq{thinpro} to the gauge-mediated case. By linearising the gauge-mediated potential, we obtained the analytical properties instead. For both potential types, we both analytically and numerically examined characteristic slopes as well as the stability of the $Q$-balls in the thin and thick-wall limits. Our main analytical results are summarised in \fig{tbl:results}.

\vspace*{10pt}

This present paper is of course related to our previous work  \cite{Tsumagari:2008bv}. The key differences are that in the present work on thin-wall $Q$-balls we are assuming the value of $\sigma_+(\omega)$ for the thin-wall limit $\omega\simeq \omega_-$ depends weakly on $\omega$ and we have replaced the assumption $\sigma(R_Q)<\sigma_-(\omega)$ by the equivalent assumption (made by Coleman) $\Uo\simeq U_{\omega_-}$ in the $Q$-ball shell region \cite{Coleman:1977py}. These in turn are related to the previous requirement that the surface tension $\tau$ depends weakly on $\omega$, which can be translated into the main assumptions: $R_Q\gg \delta, 1/\mu, \sigma_0\simeq \sigma_+$, and $\Uo\simeq U_{\omega_-}$ in the shell region. Furthermore, our analytic work agrees well with the numerical results for small curvature $\mu$ with $|K|=0.1$; however, it is not clear that our analytic framework still holds even in the case of $|K|\ll \order1$, which corresponds to a case where the potential is extremely flat, see \eq{mucurv}.

\paragraph*{\underline{\bf $Q$-balls in gravity-mediated potentials:}}

It is possible to obtain absolutely stable $Q$-matter with a small coupling constant, \eq{restbeta}, for the nonrenormalisable term in \eq{sugra}. For $|K|\not\ll \order1$, a gravity-mediated potential cannot be really considered as flat, which allows us to apply our previous results, \eqs{grvthnch}{grvthncls}, in \cite{Tsumagari:2008bv} to describe the thin-wall $Q$-ball where $\sigma_0\simeq \sigma_+$. In the thick-wall limit by reparameterising parameters in $\So$ and neglecting the nonrenormalisable term under the conditions $\beta^2 \lesssim |K| \lesssim \order{1}$, we have obtained the stationary properties of the $Q$-ball. We showed that the thick-wall $Q$-ball is classically stable, and demonstrated that under certain conditions \eq{thckabs3} it can be absolutely stable. Although this analysis is much simpler than the analysis associated with imposing a Gaussian ansatz developed in Appendix \ref{appxthick}, the former analysis assumed that the nonrenormalisable term is negligible at the beginning of the analysis. In the latter analysis, we have kept all terms in \eq{repsugra} and shown that the nonrenormalisable term is indeed negligible in the limit $\omega \gtrsim \order{m}$. Our results, \eqs{grvclscond}{grvch}, for the thick-wall $Q$-ball have recovered the previous results obtained in \cite{Enqvist:1998en, Morris:1978ca} without any contradictions for classical stability conditions as opposed to the case of using a Gaussian ansatz in a general polynomial potential in which we showed that the ansatz led to a contradiction and corrected it by introducing a physically motivated ansatz \cite{Tsumagari:2008bv}. This is because the Gaussian ansatz, \eq{testgauss}, becomes the exact solution, \eq{gauss}, in the Gravity mediated potential in the limit $\omega \gtrsim \order{m}$ where the nonrenormalisable term is negligible. In \figs{fig:grvst}{fig:grvabs} the analytical lines agree well with the corresponding numerical plots in both the thin-wall and thick-wall limits. Under our numerical parameter sets, the $Q$-balls in DVP are both classically and absolutely stable up to $\omega \lesssim m$, while all of the $Q$-balls in NDVP are absolutely unstable because of our choice, $\omega_-=m$. We believe that an absolutely stable $Q$-matter exists in NDVP when we take $\omega_-<m$. Since the $Q$-balls in both potential types are always classically stable, as can be seen in the top two panels of \fig{fig:grvabs} except for the case of $1D$ $Q$-balls in the thin-wall limit to which our analytical work cannot be applied since it holds only for $D\ge 2$. We have also found the asymptotic profile \eq{asympro} for all possible values of $\omega$, see the top two panels in \fig{fig:grvpro}.

Our analytical estimations on the value of $\frac{\omega}{Q}\frac{dQ}{d\omega}$ do not agree well with the numerical results, because $\sigma_0\not\sim \sigma_+$. Nevertheless the other analytical properties are well fitted especially in NDVP, see bottom panels in \figs{fig:grvst}{fig:grvabs}. The DVP in \eq{sugra} for small $|K|$ is extremely flat as the gauge-mediated potential in \eq{potgauge} where both of the potentials have $\omega_-\simeq 0$. Notice that the asymptotic profile for the former case has a Gaussian tail, while the latter profile is determined by the usual quadratic mass term, see \eqs{asympro}{gauasym}. By assuming that the shell effects are much smaller than the core effects in the thin-wall limit, the difference of the tails can be negligible. Indeed, we can see the thin-wall numerical lines for both the classical stability and the characteristic slope look qualitatively and quantitatively similar to each other, as can be seen in both the top/bottom left panels of \fig{fig:grvabs} and the panels of \fig{fig:gauabs}. Notice that the spikes of energy density in the gauge-mediated potential cannot be seen even though $\omega_-\simeq 0$, see \fig{fig:gaupro}.

Furthermore, we know that the potential $U_{grav}$ can be approximated by $\half m^2 M^{2|K|}\sigma^{2-2|K|}$ for small $|K| \ll \order1$, then the potential in \eq{sugra} looks similar to the confinement model in \cite{Simonov:1979rd, Mathieu:1987mr}. By neglecting the nonrenormalisable terms in the thick-wall limit, we can easily obtain the characteristic slope, $\gamma = \frac{2+|K|(D-1)}{2+|K|(D-2)}\simeq 1$, \cite{Dine:2003ax} by following the same technique as in \eq{leg}, which does not depend on $\omega$ but does depend on $D$ and $|K|$. It follows that $E_Q\propto Q^{1/\gamma}$ from \eq{leg2}. This result is obviously worse than our main results in \eqs{grvqeq}{grvch}, see bottom two panels in \fig{fig:grvst}, because we know that the Gaussian ansatz \eq{testgauss} can be the exact solution \eq{gauss} for $U=U_{grav}$; thus, it is not so powerful to approximate $U_{grav}$ by $\half m^2 M^{2|K|}\sigma^{2-2|K|}$ for small $|K|$.

\paragraph*{\underline{\bf $Q$-balls in gauge-mediated potentials:}}

For the Gauge mediated potential in \eq{potgauge}, we obtained the full analytic results in $D=1,\; 3$ over the whole range of $\omega$ using \eqs{gauR1}{gauR3}, see \figs{fig:gaueq}{fig:gauabs}. In the thin-wall limit for $\mo R,\; \omega R \gg \order{1}$, we reproduced the previously obtained results, \eq{gauthnch}, in \cite{Dvali:1997qv, MacKenzie:2001av, Asko:2002phd} and showed that they are both classically and absolutely stable in \eqs{gauthnch}{gaucls}. The one- and three-dimensional thick-wall $Q$-balls, on the other hand, are neither classically nor absolutely stable, see either \eqs{1dcls}{1dch} or \eqs{3dcls}{gauthinch}, respectively. Since the potential, \eq{potgauge}, is not differentiable everywhere, we have used the approximate potential, \eq{apprxpot}, instead in the numerical section. \figs{fig:gaueq}{fig:gauabs} show that the numerical results agree with the analytical results in the thin-wall limit. The numerical data near the thick-wall limit and/or in the $1D$ case differ from the analytic lines since the profiles are computed in the region where the two potentials between \eq{apprxpot} and \eq{potgauge} are different, see \fig{fig:gaupot}. This differences come from the artifact of our approximated smooth potential \eq{apprxpot} against the generalised gauge-mediated potential \eq{potgauge}.

\paragraph*{\underline{\bf The $3D$ $Q$-balls:}}
Although we have shown $Q$-ball results for an arbitrary number of spatial dimensions $D$, only three-dimensional cases are phenomenologically interesting. $Q$-balls in flat potentials give the proportional relation $E_Q \propto Q^{1/\gamma}$, where $\gamma$ generally depends on $D$. The actual values of $1/\gamma$ for three-dimensional thin-wall $Q$-balls are $\frac{4}{5},\; 1,$ and $\frac{3}{4}$ in DVP, NDVP of gravity-mediated potentials and in gauge-mediated potentials respectively. It implies that the gauge-mediated $Q$-balls would be formed in the most energetically compact state for a large charge $Q$, so it is likely that such formed $Q$-balls would have survived any possible decay processes and thermal evaporation until the present day, and possibly become a dark matter candidate  \cite{Kusenko:1997si}.

\paragraph*{\underline{\bf Dynamics and cosmological applications:}}

The dynamics of a pair of one-dimensional $Q$-balls has been recently analysed using momentum flux \cite{Bowcock:2008dn}. 
For a large separation between the $Q$-balls, the profiles develop the usual exponential tail, $e^{-\mo r}$, in general polynomial potentials and  in \cite{Bowcock:2008dn} the authors showed that there was a solitonic force between them. Profiles in the gravity-mediated models and other confinement models, however, have different asymptotic tails, which may affect the detailed dynamics and the $Q$-ball formation \cite{Kasuya:2000wx, Multamaki:2000qb, Multamaki:2001az, Kusenko:2009cv}.

In a cosmological setting (thermal background), SUSY $Q$-balls are generally unstable via evaporation, diffusion, dissociation, and/or decay into todays baryons and lightest supersymmetric particles, if the AD field couples to the thermal plasma, which are decay products from inflaton, and/or if the field possesses a lepton number for the MSSM flat directions \cite{Enqvist:1997si, Enqvist:1998en}. Following our detailed analytical and numerical analyses of both gravity-mediated and gauge-mediated $Q$-balls, it is clear that this whole area of dynamics and cosmological implications of these $Q$-balls deserves further analyses.

\begin{figure}[!ht]
  \def\@captype{table}
  \begin{minipage}[t]{\textwidth}
   \begin{center}
      \begin{tabular}{|c||c|c|c||c|c|}
	\hline
	Model & \multicolumn{3}{|c||}{Gravity mediated} & \multicolumn{2}{|c|}{Gauge mediated}\\
    	\hline
	$Q$-ball type & \multicolumn{2}{|c|}{Thin-wall} & Thick-wall & Thin-wall & Thick-wall \\ \hline
	Conditions & \multicolumn{2}{|c|}{$\blacktriangle$} & $\beta^2 \lesssim |K|\lesssim \order{1}$ & None & $D=1,3, ...$\\
	\hline
	Assumptions & \multicolumn{2}{|c|}{$R_Q \gg \delta, 1/\mu;\; \sigma_0 \simeq \sigma_+$ and $\Uo\simeq U_{\omega_-}$ in shell} & None & $R\gg 1/\mo, 1/\omega$ & None\\
	\hline
	Potential type & DVPs & NDVPs & Both & \multicolumn{2}{|c|}{NDVPs} \\
    	\hline
	& & & & & \\
	$1/\gamma$ & $\frac{2D-1}{2(D-1)}$ & 1 & 1 & $\frac{D}{D+1}$ & 1 \\
	Absolute stability & $\bigcirc$ & $\bigtriangleup$ & $\bigtriangleup$ & $\bigcirc$ & $\times$\\
	Classical stability & $\bigcirc$ & $\bigcirc$ & $\bigtriangleup$ & $\bigcirc$ & $\times$\\
\hline
      \end{tabular}
	\end{center}
	\tblcaption{Key analytical results. Recall that the $\omega$-independent characteristic slope $\gamma \equiv E_Q/\omega Q$ leads to the proportionality relation $E_Q\propto Q^{1/\gamma}$. The symbols, $\bigcirc,\; \times,\; \bigtriangleup$, indicate that $Q$-balls are stable, unstable, or can be stable with conditions, respectively. The symbol, $\blacktriangle$, means that we may need the condition $|K|\not\ll \order{1}$.}
    \label{tbl:results}
  \end{minipage}
\end{figure}

\section*{Acknowledgement}
The work of M.I.T. is supported by Nottingham University studentships. E.J.C. is grateful to the Royal Society for financial support. M.I.T. would like to thank N. Bevis, J. McDonald, P. Saffin, and O. Seto for useful discussions and correspondence.

\begin{appendix}

\section{An exact solution}\label{exactsol}

In this appendix we will show that a Gaussian profile is an exact solution of the $Q$-ball equation in \eq{QBeq} with $\Uo=U_{grav}-\half \omega^2\sigma^2$ in \eq{sugra}. Notice that the potential $U_{grav}$ becomes negative for $e^{1/2|K|} M < \sigma$, hence the system is not bounded from below. The additional contribution from the nonrenormalisable term $U_{NR}$ compensates the negative term and supports the existence of $Q$-balls in the system. Although the Gaussian exact solution is no longer a solution for the full potential $U_{grav}+U_{NR}$ in \eq{sugra}, the solution we will obtain here provides hints in suggesting a reasonable ansatz for the thick-wall $Q$-ball as we will see later. 

Let us consider the following Gaussian profile:
\be{gauss}
\sigma_{sol}(r)=\rho_{\omega}\ \exp\bset{-\frac{|K|m^2 r^2}{2}},
\ee
where we will see that $m,\; M,$ and $|K|$ are the same parameters as in \eq{sugra} and $\rho_{\omega}$ will be shortly determined in terms of the underlying parameters. By substituting \eq{gauss} into the left-hand side of 
\eq{QBeq} it leads to 
\be{potgauss}
U_{grav}=\frac{m^2}{2}\sigma^2\bset{1-|K|\ln\bset{\frac{\sigma}{M}}^2}
\ee
and 
\be{rhom}
\rho_{\omega}= M \exp\bset{\frac{D-1}{2}+ \frac{\mo^2}{2 |K| m^2}},
\ee
where we set the integration constant as zero. Recall $\mo^2 \equiv m^2-\omega^2$. Note that the constant $M$ has the same mass dimension, $(D-1)/2$, as $\sigma$ so that the only physical case is $D=3$. The profile, \eq{gauss}, is an exact solution for $U_{grav}$ with the ``core'' radius $R_Q=\sqrt{2/m^2|K|}$ \cite{Enqvist:1998en}, which is very large compared with $m^{-1}$ for small $|K| \ll \order{1}$, and satisfies the boundary conditions for $Q$-balls, namely $\sigma^{\p}(0)=0=\sigma(\infty)=\sigma^\p(\infty)$ \cite{Tsumagari:2008bv}. In the extreme limit $\omega \gg m $, we obtain $\rho_{\omega}\to 0$ {for $|K|\lesssim \order{1}$} which implies $\sigma_0\equiv \sigma(0) \to 0$. For large $\sigma$, the potential becomes asymptotically flat, tending towards an infinite negative value. By adding the nonrenormalisable term $U_{NR}$, the potential $U_{grav}$ is lifted for large $\sigma$ in \eq{sugra}, then the full potential $U_{grav}+U_{NR}$ is bounded from below, see Sec. \ref{exstQmat}. We can see the ansatz given in \cite{Enqvist:1998en} corresponds to the case where $\rho_\omega \simeq M$, which is valid only for $|K| \ll \order{1}$ and $\omega \simeq m$, see \eq{rhom}.

\section{Thick-wall $Q$-ball with a Gaussian ansatz}\label{appxthick}

In this appendix, we will investigate the thick-wall $Q$-ball in gravity-mediated models by introducing a Gaussian ansatz and keeping all terms in \eq{repsugra} as opposed to the analysis in Sec.\ref{thickgrav}. By using this profile we can perform the Gaussian integrations, and will obtain the generalised results of \eqs{grvqeq}{grvclscond} in Sec. \ref{thickgrav}. The test profile  for the case, $\omega \gtrsim \order{m}$, coincides with the solution $\sigma_{sol}$ in \eq{gauss}, which implies that the nonrenormalisable term $U_{NR}$ in \eq{sugra} is negligible.

To recap, the notation we have adopted in \eq{repsugra} is $\tisig=\sigma/M,\; \tiom=\omega/m$, $\beta^2$ is defined in \eq{beta2} and we are considering the case of $n>2$. To begin with we introduce a Gaussian ansatz inspired by \eq{gauss} for the potential \eq{repsugra}
\be{testgauss}
\tisig(r)=\lamom \exp(-\alom^2 r^2/2),
\ee
where $\tisig_0 \equiv \tisig(0)=\lamom={\rm finite}$, and $\lamom,\; \alom$ will be functions of $\omega$ implicitly. $\lamom$ should not be confused with the coupling constant $\lambda$ in \eq{sugra}. Both $\lamom$ and $\alom$ can be determined by extremising the Euclidean action $\So$; hence the actual free parameter here will be only $\omega$. It is crucial to note that $\lamom$ cannot be infinite in the thick-wall limit since we know that $\lamom$ is finite and tending to 0. If the nonrenormalisable term $U_{NR}$ is negligible, we can expect $\lamom \sim \rho_\omega/M \sim \tisig_-(\omega)$ and $\alom^2 \sim |K|m^2$ due to \eq{gauss}, which implies that the ``core'' radius $R_Q$ of the thick-wall $Q$-ball is $R_Q\sim\sqrt{2/m^2|K|}$. For the extreme thick-wall limit $\omega \gg m$, we shall also confirm $\lamom \to 0$, which means $\tisig_0 \to 0$.
 
By substituting \eq{testgauss} into \eq{euc} with the potential \eq{repsugra}, we obtain $Q$ and $\So$ using the following Gaussian integrations:
$\Omega_{D-1}\int^\infty_0 dr r^{D-1} e^{-k r^2}=\bset{\frac{\pi}{k}}^{D/2}$ for real $k$ where $\Omega_{D-1}\equiv \frac{2\pi^{D/2}}{\Gamma(D/2)}$. Thus,
\bea{qapx}
Q&=& M^2 \pi^{D/2} \omega \lamom^2 \alom^{-D}, \\
\label{soapx}\So &=& M^2 \pi^{D/2} \alom^{-D} \sbset{A(\alom,\; \lamom) + B(\omega,\; \lamom) + C(\lamom)},\\
\textrm{where} \hspace*{10pt} A(\alom,\; \lamom)&\equiv& \frac{D\lamom^2 }{4}(\alom^2+|K|m^2),\\
\label{ABC}  B(\omega,\; \lamom)&\equiv& \frac{m^2\lamom^2}{2}\bset{1-\frac{\omega^2}{m^2}- 2|K|\ln\lamom},\; C(\lamom)=m^2 \beta^2 \lamom^{n}\bset{\frac{2}{n}}^{D/2}.
\eea
Notice that $A(\alom,\; \lamom)$ comes from the gradient term and the logarithmic term in $\So$ and depends on both $\alom$ and $\lamom$. Similarly, $B(\omega,\; \lamom)$ is given by the quadratic term in the potential \eq{repsugra} and depends both on $\lamom$ and explicitly on $\omega$, whereas $C(\lamom)$ arises simply from the nonrenormalisable term in the potential. An alternative (but in this case more complicated) approach to obtain $Q$ would be the use of Legendre transformations in \eq{leg}.

By extremising $\So$ in terms of the two free parameters $\alom$ and $\lamom$: 
\be{ext}
\frac{\partial \So}{\partial \alom}=0,\hspace*{15pt} \frac{\partial \So}{\partial \lamom}=0,
\ee
we obtain 
\be{abc}
A+B+C=\frac{\lamom^2 \alom^2}{2},\hspace*{15pt} A+B+\frac{nC}{2}=\frac{m^2\lamom^2|K|}{2},
\ee
which implies that
\be{al2}
\frac{\alom^2}{m^2} = |K|-(n-2)\beta^2 \lamom^{n-2} \bset{\frac{2}{n}}^{D/2} \ge 0,
\ee
where we have eliminated the $A+B$ terms in the two expressions of \eq{abc}. Using \eq{al2} and the second expression of \eq{abc}, we obtain the relations between $\omega$ and $\lamom$
\bea{lam2}
\frac{\omega^2}{m^2}&=& 1 + |K|\bset{D-1- 2\ln\lamom} + \frac{2(n+D)-nD}{2}\beta^2 \lamom^{n-2}\bset{\frac{2}{n}}^{D/2},\\
 \label{lam4}   &\sim&
    \left\{
    \begin{array}{ll}
    1+|K|(D-1-2\ln\lamom) &\; \textrm{for}\ |K| \sim \order{1},  \\
    1-2|K|\ln\lamom  &\; \textrm{for}\ |K|<\order{1},
    \end{array}
    \right.\\
\label{lam3} \frac{d \lamom}{d\omega}&=& -\frac{\lamom \omega}{|K|m^2 F}\sim - \frac{\lamom \omega}{|K|m^2} <0,
\eea
where we have differentiated \eq{lam2} with respect to $\omega$ to obtain \eq{lam3} and have defined $F$ as $F\equiv 1 - (n-2)\frac{2(n+D)-nD}{4}\frac{\beta^2}{|K|} \lamom^{n-2} \bset{\frac{2}{n}}^{D/2}= 1+ \frac{2(n+D)-nD}{4|K|m^2}\bset{\alom^2 -m^2|K|}$. Equations (\ref{al2}, \ref{lam2}) imply that both $\alom$ and $\lamom$ are functions of $\omega$; however, these are not solvable in closed forms unless the particular limits, which were introduced in Sec.\ref{thickgrav}, are taken, as we will now show. Comparing \eqs{lam2}{lam3} with \eqs{thckom1}{thckom3}, we can see an extra contribution of $\order{|K|}$ in \eq{lam2}, which is not present in \eq{thckom1}. This difference of $(D-1)|K|$ arises because in  calculating \eq{lam2} we have used  $\lamom$, whereas we have used $\tisig_-(\omega)$ in obtaining \eq{thckom1}, and although related they are not precisely the same. In the extreme thick-wall limit $\omega \gg m$, and from \eq{lam2} this implies $\lamom \to 0^+$ (recall from \eq{testgauss} that $\lamom$ has to remain finite). Considering the nonrenormalisable term in \eq{lam2}, the fact that $\beta^2 \lesssim |K|\lesssim \order{1}$ and $\lamom \to 0^+$ with $n>2$, implies that this term is subdominant and can be ignored. 
As long as $\lamom < \order{1}$, then $F\sim 1$ and the second relation of \eq{lam3} follows, which implies that $\lamom$ is a monotically decreasing function in terms of $\omega$. The limit $\lamom \sim \order{1}$ corresponds to $\omega \gtrsim \order{m}$, see \eq{lam2}. We will call this the  ``moderate limit'' and represent it by '$\sim$'. The other case, $\omega \gg m$ (or equivalently $\lamom \ll \order{1}$), we shall call the ``extreme limit'' and represent it by '$\to$'. Depending on the logarithmic strength of $|K|$, we can obtain \eq{lam4}, which leads to the approximated expressions for $\lamom$ and can also obtain $\alom$ from \eq{al2}
\be{predict}
    \lamom \sim
    \left\{
    \begin{array}{ll}
    \rho_\omega/M &\; \textrm{for}\ |K| \sim \order{1} \\
    \tisig_-(\omega)  &\; \textrm{for}\ |K|<\order{1}
    \end{array}
    \right.
    \to 0;\; \;
\frac{\alom^2}{m^2} \sim |K| \to |K| \; \textrm{for}\ |K| \lesssim \order{1},
\ee
where $\alom$ is independent of $\omega$ in both the ``moderate'' and ``extreme'' limits.

Using \eqs{qapx}{soapx} and \eq{abc}, we obtain the characteristic slope in both the ``moderate'' and ``extreme'' limits,
\be{grvch}
 \frac{E_Q}{\omega Q}= 1+ \frac{\alom^2}{2\omega^2} \sim 1 + \frac{m^2|K|}{2\omega^2} \to 1.
\ee
In order to show their classical stability, we shall differentiate $Q$ with respect to $\omega$ using \eqs{al2}{lam2} and \eq{lam3}:
\bea{}
\nonumber\frac{\omega}{Q}\frac{dQ}{d\omega}&=&1 - \frac{2\omega^2}{m^2|K|F} \sbset{1-\frac{D(n-2)}{4\alom^2}\bset{\alom^2-m^2|K|} },\\
\label{grvclscond2} &\sim& 1- \frac{2\omega^2}{m^2|K|} \to -\frac{2\omega^2}{m^2|K|}<0,\\
\nonumber \frac{d}{d\omega}\bset{\frac{E_Q}{Q}}&=&1-\frac{1}{2\omega^2}\sbset{\alom^2 + \frac{(n-2)\omega^2}{m^2|K|F}\bset{\alom^2-m^2|K|}},\\ 
\label{grvclscond3} &\sim& 1-\frac{m^2|K|}{2\omega^2} \to 1 >0,
\eea
where we have taken the ``moderate limit'' and ``extreme limit'' and used $\alom^2 \sim m^2 |K|,\; F=1+ \frac{2(n+D)-nD}{4|K|m^2}\bset{\alom^2 -m^2|K|}\sim 1$. The classical stability condition \eq{grvclscond2} is consistent with \eq{grvclscond3}, and is consistent with \eq{QBcls}. This is different from the result we obtained for the polynomial potentials [see Eq. (74) in  \cite{Tsumagari:2008bv}], because in that case the Gaussian ansatz does not give the exact solution unlike here in  \eq{testgauss} where it does become the exact solution \eq{gauss} in both limits. The results, \eqs{grvch}{grvclscond2} and \eq{grvclscond3}, in both the  ``moderate'' and ``extreme'' limits recover the key results, \eqs{grvqeq}{grvclscond}, and are independent of $D$; hence, the thick-wall $Q$-balls for all $D$ have similar properties. We can also see the small additional effects arising from the nonrenormalisable term in \eqs{grvclscond2}{grvclscond3}.

Let us summarise the important results we found in this appendix. By introducing a Gaussian test profile \eq{testgauss} inspired by the exact solution \eq{gauss} for $U_{grav}$, we computed the Euclidean action $\So$ and the charge $Q$ using Gaussian integrations. Then, we extremised $\So$ in terms of $\lamom$ and $\alom$ in \eq{ext}, which gave the relations of both $\lamom$ and $\alom$ as a function of $\omega$. By introducing two limits called ``moderate limit'' and ``extreme limit'', we confirmed that the ansatz, \eq{testgauss}, approaches \eq{gauss} in the ``moderate limit''. We established that the results \eqs{grvch}{grvclscond2} and \eq{grvclscond3} recovered the previous results in \eqs{grvqeq}{grvclscond} which are obtained simply by reparameterising in $\So$ and extracting the explicit $\omega$-dependence from the integral in $\So$ with $U=U_{grav}$ where the nonrenormalisable term was neglected at the beginning of the analysis by applying L'H\^{o}pital rules.

In addition, we would like to emphasise the main differences between our work and other earlier analyses in the literature \cite{Enqvist:1998en, Morris:1978ca}. The analytical framework adopted in  \cite{Morris:1978ca} is valid only for $|K|=1,\; D=3,\; n=4$. Our work has shown that this can be generalised to arbitrary integer values of $D$ and $n (>2)$ under the conditions $\beta^2 \lesssim |K|\lesssim \order{1}$, and that the thick-wall $Q$-ball can be classically stable. In Sec.\ref{thickgrav}, we also found that the thick-wall $Q$-ball may be absolutely stable under certain additional conditions, \eq{thckabs3}. Furthermore, Enqvist and McDonald in \cite{Enqvist:1998en} analytically obtained the same ``core'' size of thick-wall $Q$-balls, although they obtained a slightly different value for $E_Q/Q$ (see their Eq. (112)). The reason for this is because their ansatz assumed $\lamom \simeq 1$ in \eq{testgauss} by simply neglecting the nonrenormalisable term, which implies that the third term of $B(\omega,\; \lamom)$ and term $C(\lamom)$ in \eq{ABC} are absent. Hence, their analysis is valid for $|K| \ll \order{1}$ and $\omega \simeq m$, see \eq{rhom}. We, however, have kept all the terms in \eq{repsugra} and used a more general ansatz, which can be applied for $|K| \lesssim \order{1}$ and $\omega \gtrsim \order{m}$ with the restricted coupling constant of the nonrenormalisable term $\beta^2 \lesssim |K|$. In summary, in this appendix we have  
extensively investigated analytically  both the absolute and classical stability of $Q$-balls in \eq{grvch} and \eq{grvclscond2}.

\end{appendix}

\end{document}